\documentclass[a4paper,12pt]{scrartcl}

\usepackage{cite}
\usepackage{xspace}
\usepackage{graphicx}
\usepackage{pslatex}
\usepackage{amsmath}
\usepackage{txfonts}
\usepackage{color}
\usepackage{multirow}
\usepackage[table]{xcolor}

\graphicspath{{figs/}}

\makeatletter 
\@ifpackageloaded{hyperref}{}{\def\url#1{#1}}
\makeatother 

\def\nn{\nonumber}

 \def\mtt{\ensuremath{M_{t\bar%
      t}}\xspace} \def\muf{\ensuremath{\mu_f}\xspace}
\def\mur{\ensuremath{\mu_r}\xspace}

\def\Bt{\ensuremath{B_t}\xspace}
\def\Chel{\ensuremath{C_{\text{hel}}}\xspace}
\def\Cnn{\ensuremath{C_{\text{nn}}}\xspace}
\def\Ckk{\ensuremath{C_{\text{kk}}}\xspace}
\def\Crr{\ensuremath{C_{\text{rr}}}\xspace}
\def\Crn{\ensuremath{C_{\text{rn}}}\xspace}
\def\Cnr{\ensuremath{C_{\text{nr}}}\xspace}
\def\D{\ensuremath{D}\xspace}
\def\OCP{\ensuremath{\mathcal{O}_{CP}}\xspace}
\def\dphi{\ensuremath{\phi^*_{CP}}\xspace} \def\sign{\mbox{sign}}

\newcommand{\hk}{{\hat{\mathbf k}}} \newcommand{\hp}{\hat{\mathbf p}}
\newcommand{\hr}{\mathbf{\hat r}} \newcommand{\hn}{\mathbf{\hat n}}
\newcommand{\ha}{{\hat{\mathbf a}}}
\newcommand{\hl}{{\boldsymbol{\hat\ell}}}
\newcommand{\hb}{{\hat{\mathbf b}}} \newcommand{\St}{{\mathbf S}_t}
\newcommand{\Stbar}{{\mathbf S}_{\bar t}} \newcommand{\Stx}{{\mathbf%
S}_{\bar{t}}}
\newcommand{\R}{r}  
\newcommand{\Y}{y}  

\newcommand{\Obs}{\mathcal{O}}

\def\vev{{\varv}}

\def\Ref#1{Ref.~\cite{#1}}
\def\Eq#1{Eq.~(\ref{#1})} 
\def\Fig#1{Fig.~\ref{#1}}
\def\Figs#1{Figs.~\ref{#1}} 
\def\Sec#1{Sec.~\ref{#1}}

\def\Tab#1{Tab.~\ref{#1}}

\def\CP{$CP$\xspace}

\newcommand{\ttbar}{\ensuremath{t{\bar t}}\xspace}

\newcommand{\hlp}{\hat{\boldsymbol{\ell}}_+}
\newcommand{\hlm}{\hat{\boldsymbol{\ell}}_-}
\newcommand{\hlpm}{\hat{\boldsymbol{\ell}}_{\pm}}

\begin{document}

%

\begin{titlepage}

  \begin{flushright}
    HU-EP-17/04\\ TTK-17-03 \\
  \end{flushright}
  \vspace{0.01cm}
  
  \begin{center}
    {\LARGE \textbf{Production of heavy Higgs bosons\\ and decay into top
      quarks at the LHC.\\[0.3em] II: Top-quark polarization and spin
      correlation effects}}\\
    \vspace{1.5cm}
    {\textbf{ W. Bernreuther}}\,$^{a,}$\footnote{\texttt{
    breuther@physik.rwth-aachen.de}}, {\textbf{
    P. Galler}}\,$^{b,}$\footnote{\texttt{ galler@physik.hu-berlin.de}},
    {\textbf{Z.-G. Si}}\,$^{c,}$\footnote{\texttt{ zgsi@sdu.edu.cn}} {\textbf{ and
    P. Uwer}}\,$^{b,}$\footnote{\texttt{ peter.uwer@physik.hu-berlin.de}}
    \par\vspace{1cm}
    $^a$Institut f\"ur Theoretische Teilchenphysik und Kosmologie,
    RWTH Aachen University, \\ 52056 Aachen, Germany\\ $^b$Institut
    f\"ur Physik, Humboldt-Universit\"at zu Berlin, 12489 Berlin,
    Germany\\ $^c$School of Physics, Shandong University, Jinan,
    Shandong 250100, China
    \par\vspace{1cm}
    \textbf{Abstract}\\
    \parbox[t]{\textwidth}{%
    We analyze, within several parameter scenarios of type-II
    two-Higgs doublet extensions of the standard model, the impact
    of heavy neutral Higgs-boson resonances on top-quark pair
    production and their subsequent decay to dileptonic final
    states at the LHC (13 TeV).  In particular, we investigate the
    effects of heavy Higgs bosons on top-spin observables, that is,
    the longitudinal top-quark polarization and top-quark spin
    correlations.  We take into account NLO QCD as well as weak
    interaction corrections and show that top-spin observables, if
    evaluated in judiciously chosen top-quark pair invariant mass
    bins, can significantly enhance the sensitivity to heavy Higgs
    resonances in top-quark pair events.
    }
  \end{center}
  \vspace*{0.7cm}
 

\end{titlepage}

%

\setcounter{footnote}{0}
\renewcommand{\thefootnote}{\arabic{footnote}} \setcounter{page}{1}

%

\section{Introduction} 
\label{sec:intro}

One of the central issues of present and future research at the Large
Hadron Collider (LHC) is the search for new, in particular heavy
(pseudo-)scalar bosons with masses below or around 1 TeV. The
existence of additional spin-zero resonances besides the 125 GeV Higgs
resonance \cite{Aad:2012tfa,Chatrchyan:2012xdj} is theoretically well
motivated \cite{Djouadi:2005gj,Branco:2011iw}. Experimental searches
were negative so far (cf. for instance \cite{CanICHEP2016}), but are
by far not exhaustive and will continue with increased effort at the
run II of the LHC at 13 TeV center-of-mass energy.

A possibility that has received increased experimental attention
recently
\cite{Chatrchyan:2013lca,Aad:2015fna,Khachatryan:2015sma,ATLAS:2016pyq}
is the existence of one or several neutral Higgs bosons with masses
above the top-antitop quark ($\ttbar$) production threshold that
strongly couple to top quarks but have suppressed couplings to the
weak gauge bosons and to $d$-type quarks and charged leptons.  The
search for such a Higgs boson (or Higgs bosons) in $\ttbar$ production
is difficult because one expects that its line shape in the $\ttbar$
invariant mass spectrum is not a distinctive resonance bump but is
significantly distorted which is caused by the interference of the
signal and the nonresonant $\ttbar$ background amplitudes. As far as
experimental investigations are concerned, this interference effect
was taken into account only in the recent data analysis
\cite{ATLAS:2016pyq} of the ATLAS experiment.

Future experimental explorations of this search channel aiming at an
increased sensitivity to such type of Higgs bosons require for their
interpretation theoretical investigations beyond leading-order (LO)
QCD. In \cite{Bernreuther:2015fts} we have investigated, within the
type-II two-Higgs-doublet extension (2HDM) of the standard model (SM),
the production of heavy neutral Higgs bosons and their decay to
$\ttbar$ pairs including signal-background interference at
next-to-leading-order (NLO) in the QCD coupling. The NLO QCD
corrections to Higgs production and signal-background interference
were computed in the heavy top-quark mass limit with an effective
K-factor rescaling. The primary aim of this exploration was to analyze
how the QCD corrections affect the distortions of the $\ttbar$
invariant mass spectrum and of other distributions in the resonance
region.  Another NLO QCD analysis of heavy Higgs-boson production in
the $\ttbar$ channel including interference with the $\ttbar$
background was recently presented in \cite{Hespel:2016qaf}.

In this paper we extend our investigations of
\cite{Bernreuther:2015fts} by taking the $t$ and $\bar t$ polarization
and $\ttbar$ spin correlations fully into account at NLO QCD. We
consider, again within the type-II 2HDM, the production of heavy
neutral Higgs bosons and their decay to $\ttbar$ with subsequent decay
of $\ttbar$ to dileptonic final states, including signal-background
interference and the irreducible non-resonant $\ttbar$ background at
NLO QCD. In our computation of the nonresonant $\ttbar$ background the
mixed QCD-weak interaction corrections are also included. Our goal is
to analyze the sensitivity of top-spin observables to heavy Higgs
resonances. We explore lepton angular distributions and dileptonic
angular correlations \cite{Bernreuther:2004jv,Bernreuther:2015yna}
that are induced by $t$ and $\bar t$ polarizations and $\ttbar$ spin
correlations. We compute these observables in appropriately chosen
$\ttbar$ invariant mass windows in the standard model and in the
presence of heavy Higgs resonances. Heavy Higgs-boson effects on
$\ttbar$ spin correlations were previously investigated at LO QCD in
\cite{Bernreuther:1993hq,Bernreuther:1997gs,Bernreuther:1998qv,Baumgart:2011wk,
Barger:2011pu}.

The paper is organized as follows. In \Sec{sec:2hdm} we briefly
recapitulate the salient features of the type-II 2HDM with and without
Higgs-sector \CP violation that are relevant for our analysis and
define four parameter scenarios. In \Sec{sec:obs} we introduce spin
dependent observables.  In \Sec{sec:res} we show in analogy to what
has been done in \Ref{Bernreuther:2015fts} results for the invariant
mass distribution of the $\ttbar$ pair at the level of the
intermediate top quarks. These results are then compared, as far as
the sensitivity to heavy Higgs bosons is concerned, with predictions
for spin dependent observables in dileptonic $\ttbar$ events. All
results are shown for the LHC operating at 13 TeV and are presented
both for selected \mtt bins and inclusively in \mtt.  We conclude in
\Sec{sec:summary}.

%

\section{Parameter scenarios for the type-II two-Higgs-doublet model}
\label{sec:2hdm}

For definiteness we choose, as in \cite{Bernreuther:2015fts}, the
type-II two-Higgs-doublet model for describing both the 125 GeV Higgs
resonance and additional neutral heavy Higgs bosons in a consistent
field-theoretic and phenomenologically viable framework. We recall
that in 2HDMs the SM field content is extended by an additional Higgs
doublet. In the type-II 2HDM the Higgs doublet $\Phi_1$ is coupled to
right-chiral down-type quarks and charged leptons, while $\Phi_2$ is
coupled to right-chiral up-type quarks only. By construction,
flavor-changing neutral currents are absent at tree level.\\ Because
we are interested in heavy Higgs bosons of different \CP nature we
consider two variants of the type-II 2HDM: one where the tree-level
Higgs potential $V(\Phi_1,\Phi_2)$ is \CP-violating and, as a special
case, the model where $V(\Phi_1,\Phi_2)$ is \CP-invariant. We denote
the three physical neutral Higgs mass eigenstates by $\phi_j$,
$j=1,2,3$. Using the unitary gauge they are related to the two
\CP-even and the \CP-odd states $\varphi_{1,2}$ and $A$, respectively,
by an orthogonal transformation:
\begin{equation}\label{eq:ortho}   
  (\phi_1,\phi_2,\phi_3)^T= R(\varphi_1,\varphi_2, A)^T \, ,
\end{equation}
where $R$ is a real orthogonal matrix that is parametrized by three
mixing angles $\alpha_i$.  We use the parametrization of $R$ given in
\cite{Bernreuther:2015fts}. The masses of the three neutral Higgs
bosons and of the charged Higgs boson $H^\pm$ of the model are denoted
by $m_j$ $(j=1,2,3)$ and by $m_+$, respectively. The parameter
$\tan\beta=\vev_2/\vev_1$ is the ratio of the vacuum expectation
values of the two Higgs doublet fields with
$\vev=\sqrt{\vev_1^2+\vev_2^2}=246$ GeV. In the case of Higgs sector
\CP violation we choose, as in \cite{Bernreuther:2015fts},
\begin{equation} \label{eq:setP}
  m_1, \, m_2, \, m_3, \, m_+ ,\, \alpha_1 , \, \alpha_2 , \, \alpha_3 ,
  \, \tan\beta , \, \vev \, ,
\end{equation}
to belong to the set of independent parameters of the model, while in
the case where $V(\Phi_1,\Phi_2)$ is \CP-invariant we choose
\begin{equation} \label{eq:setPrime}
  m_1, \, m_2, \, m_3, \, m_+, \, \alpha_1,\, \tan\beta,\, \vev \, .
\end{equation} 
In this case the matrix $R$ is block-diagonal with
$R_{13}=R_{23}=R_{31}=R_{32} = 0$ and $R_{33} = 1$, and the neutral
Higgs mass eigenstates consist of two \CP-even and a \CP-odd state
which are often denoted by $\phi_1=h$, $\phi_2=H$, and $\phi_3=A$. In
the more general case of Higgs sector \CP violation the $\phi_j$ are
\CP mixtures.\\ The interactions of the $\phi_j$ with quarks and
charged leptons $f=q,\ell$ and with weak gauge boson pairs are given
by
\begin{equation} \label{eq:LfWZ}
  {\cal L}_1= -\frac{m_f}{\vev}\left( a_{jf} {\bar f}f - b_{jf}{\bar f}
    i\gamma_5 f\right) \phi_j + f_{jVV} \phi_j\left(\frac{2
      m_W^2}{\vev}W^-_\mu W^{+\mu} + \frac{m_Z^2}{\vev} Z_\mu Z^\mu \right)
  \, ,
\end{equation}
where a sum over $f$ and $j=1,2,3$ is understood. The reduced scalar
and pseudoscalar Yukawa couplings $a_{jf}$ and $b_{jf}$ and the
reduced couplings $f_{jVV}$ depend on the values of $\tan\beta$ and on
the elements $R_{ij}$ of the Higgs mixing matrix. They are listed in
\Tab{tab:LfWZ} for the type-II model.  We recall that the reduced
gauge-boson couplings obey the sum rule $\sum_j f^2_{jVV}=1$. For
computing the widths of the heavy Higgs bosons below we need also the
triple Higgs and the $Z\phi_i\phi_j$ interactions.  They are given,
for instance, in \cite{Bernreuther:2015fts,Chen:2015gaa,
Mellein:540471}.
\vspace{2mm}
\begin{table}[htbp]
\begin{center}
  \caption{Reduced couplings to quarks, leptons and gauge bosons of the neutral
    Higgs bosons $\phi_j$ of the type-II 2HDM defined in \Eq{eq:LfWZ}. The
    labels $t,b$, and $\tau$ refer to $u$-type, $d$-type quarks, and
    charged leptons.}  \vspace{1mm}
  \begin{tabular}{cccc|c}
    $a_{jt}$ & $a_{jb} = a_{j\tau}$ & $b_{jt}$ & $b_{jb} = b_{j\tau}$ &
    $f_{jVV}$\\ \hline \hline $R_{j2}/\sin\beta$ & $R_{j1}/\cos\beta$ &
    $R_{j3}\cot\beta $ & $R_{j3}\tan\beta$ & $\cos\beta R_{j1}+ \sin\beta
    R_{j2}$
  \end{tabular}
  \label{tab:LfWZ}
\end{center}
\end{table}%
We identify the 125 GeV Higgs resonance with $\phi_1$ and assume that
the mass of both $\phi_2$ and $\phi_3$ is larger than twice the
top-quark mass, $m_{2,3} > 2 m_t$. Moreover, we assume that the mass
of the charged Higgs boson $H^+$ is of the order of $\mbox{max}(m_2,
m_3)$, so that the two-body decays $\phi_{2,3} \to W^\pm H^\mp$ cannot
take place. The ATLAS and CMS results \cite{Khachatryan:2016vau} on
the 125 GeV Higgs boson imply that its interactions with the
third-generation fermions and gauge bosons are SM-like.  This
constraint is taken into account in the 2HDM parameter scenarios
defined below.\\ As already emphasized in the introduction our aim is
to investigate the sensitivity of top-spin observables to heavy
Higgs-bosons and, in particular, whether suitable observables allow to
discriminate between a scalar, pseudoscalar, and a \CP mixture. For
this purpose we choose three parameter sets, which we call set 1a, 1b,
and 1c, where the masses of the two heavy neutral Higgs bosons are put
to 400 GeV and 900 GeV. Sets 1a and 1b are associated with a
\CP-invariant Higgs potential. In set 1a we assign the quantum numbers
\CP=+1 and \CP=-1 to the 400 GeV and 900 GeV Higgs boson,
respectively, and vice versa in set 1b. Set 1c is associated with a
\CP-violating Higgs potential and the neutral Higgs bosons are chosen
to be \CP mixtures. In addition we investigate the case where the two
heavy neutral Higgs bosons are nearly degenerate with masses that are
substantially larger than $2 m_t$. This is exemplified with a
parameter set called set 2 below.\\ We determine the total widths of
the heavy neutral Higgs bosons $\phi_2$ and $\phi_3$ for the four
parameter sets specified below, by computing the sum of the largest
two-body decay rates. We include the NLO QCD corrections to the
partial decay rates of $\phi_j\to q{\bar q}$ and $\phi_j\to g g$ using
the formulas of \cite{Braaten:1980yq,Drees:1990dq} and
\cite{Spira:1995rr}, respectively.
In these computations we use the input parameters and the scale choice
as described in \Sec{subsec:setup}.  We checked, where possible, our
results with the computer codes of \cite{Djouadi:1997yw} and
\cite{Eriksson:2009ws}.  For the parameter sets below, the total width
$\Gamma_1$ of $\phi_1(125~\mbox{GeV})$ is of the order of 4 MeV. It
plays no role in the computations of \Sec{sec:res}.
%

\subsubsection*{Scenarios 1a and 1b}

We consider the type-II model with a \CP-conserving Higgs
potential. We choose both for the parameter set 1a and 1b:
\begin{equation}
\label{eq:angle-sc1ab}
\tan\beta=1 \;,\quad \alpha_1 = \beta \; ,\quad \alpha_2 = \alpha_3 =
0 \, , \quad m_1 = 125 \mbox{\,GeV}\;, \quad m_{+} > 820 \mbox{\,GeV} \; .
\end{equation}
With this choice of the Higgs mixing angles and our convention for $R$
\cite{Bernreuther:2015fts} the states $\phi_1$ and $\phi_2$ are
\CP-even while $\phi_3$ is \CP-odd, as can be seen from the resulting
Yukawa couplings given in \Tab{tab:Yuk1ab2}. The masses of $\phi_2$
and $\phi_3$ are set to the values
\begin{align}\label{eq:mass-sc1ab}
\text{scenario 1a:} & \quad m_2 = 400 \mbox{\,GeV}\;,\quad m_3 = 900
\mbox{\,GeV}\; , \nn\\ \text{scenario 1b:} & \quad m_2 = 900
\mbox{\,GeV}\;,\quad m_3 = 400 \mbox{\,GeV}\; .
\end{align}
Thus in scenario 1a the lighter of the two heavy states is chosen to
be a pure scalar while in scenario 1b it is a pseudoscalar.\\ We
compute the largest two-body decay rates of $\phi_2$ and $\phi_3$ and
determine their total widths by adding up these rates.  The results
are listed in \Tab{tab:res_widths_sc1a} and \Tab{tab:res_widths_sc1b}
for scenario 1a and 1b, respectively. The uncertainties result from
varying the renormalization scale as described below \Eq{eq:rensc-var}. The
partial decay widths of $\phi_j\to f{\bar f}$ $(f\neq t),$ $\phi_j\to
\gamma\gamma$, and
$\phi_j\to Z\gamma$ are a few $\times 10^{-3}$ GeV or smaller%
\footnote{The partial decay width of $\phi_j\to \gamma\gamma$
  $(j=2,3)$ in these parameter scenarios and in the scenarios 1c and 2
  below is $\lesssim 3 \times 10^{-4}$ GeV.}
and are neglected in the total widths $\Gamma_2$,
$\Gamma_3$. Moreover, to lowest order in the non-QCD couplings the
partial decay rates for $\phi_i\rightarrow VV$, $\phi_i\rightarrow
\phi_1 Z$ and $\phi_i\rightarrow \phi_1\phi_1$ $(i=2,3)$ are zero for
our choice of parameters.
\begin{table}[htbp]
  \centering
  \caption{Values of the reduced couplings to fermions and  vector bosons $V=W,Z$
    of the neutral Higgs bosons $\phi_j$ in scenarios 1a, 1b and 2.}\vspace{3mm}
  {\renewcommand{\arraystretch}{1.2} \renewcommand{\tabcolsep}{0.2cm}
    \begin{tabular}{lccccc}
      \hline \hline & $a_{jt}$ & $a_{jb}=a_{j\tau}$ & $b_{jt}$ &
      $b_{jb}=b_{j\tau}$ & $f_{jVV}$\\ \hline $\phi_1$ & 1 & 1 & 0 & 0 & 1\\
      $\phi_2$ & 1 & -1 & 0 & 0 & 0\\ $\phi_3$ & 0 & 0 & 1 & 1 & 0\\ \hline
      \hline
    \end{tabular}}
  \label{tab:Yuk1ab2}
\end{table}%
\vspace{2mm}
\begin{table}[htbp]
\begin{center}
  \caption{Dominant partial decay widths and the total width of the \CP-even and
    \CP-odd Higgs boson $\phi_2$ and $\phi_3$, respectively, in scenario
    1a. }  \vspace{3mm}
  {\renewcommand{\arraystretch}{1.5} \renewcommand{\tabcolsep}{0.2cm}
    \begin{tabular}{lll}
      \hline\hline decay mode & $\Gamma_2$ [GeV] & $\Gamma_3$ [GeV]\\ \hline
      $\phi_i\rightarrow t\bar{t}$ & $3.97^{+0.10}_{-0.08}$ &
      $45.43^{-0.36}_{+0.30}$\\ $\phi_i\rightarrow \phi_2Z$ & 0 & 116.85\\
      $\phi_i\rightarrow gg$ & $0.017^{+0.003}_{-0.002}$ &
      $0.107^{-0.0003}_{-0.0036}$\\[1mm] \hline total &
      $3.99^{+0.10}_{-0.08}$ & $162.39^{-0.36}_{+0.30}$\\ \hline\hline
    \end{tabular}}
  \label{tab:res_widths_sc1a}
\end{center}
\end{table}%
\vspace{2mm}
\begin{table}[htbp]
\begin{center}
\caption{Dominant partial decay widths and the total width of the \CP-odd and
\CP-even Higgs boson $\phi_2$ and $\phi_3$, respectively, in scenario
1b.}
\vspace{3mm} {\renewcommand{\arraystretch}{1.5}
\renewcommand{\tabcolsep}{0.2cm}
\begin{tabular}{lll}
\hline\hline decay mode & $\Gamma_2$ [GeV] & $\Gamma_3$ [GeV]\\ \hline
 $\phi_i\rightarrow t\bar{t}$ & $39.85^{-0.20}_{+0.17}$ &
 $15.09^{+0.31}_{-0.26}$\\ $\phi_i\rightarrow \phi_3Z$ & 116.85 & 0\\
 $\phi_i\rightarrow gg$ & $0.068^{+0.004}_{-0.005}$ &
 $0.051^{+0.009}_{-0.007}$\\[1mm] \hline total &
 $156.76^{-0.19}_{+0.16}$ & $15.14^{+0.32}_{-0.27}$\\ \hline\hline
\end{tabular}}
\label{tab:res_widths_sc1b}
\end{center}
\end{table}%
 
%

\subsubsection*{Scenario 1c}

Here we consider the type-II model with a \CP-violating Higgs
potential and choose
\begin{equation} \label{eq:angle-sc1c}
\tan\beta=1 \;,\quad \alpha_1 = \beta \; ,\quad \alpha_2 =
\frac{\pi}{15}\;, \quad \alpha_3 = \frac{\pi}{4}\, , \quad m_1 = 125
\mbox{\,GeV}\;, \quad m_{+} > 820 \mbox{\,GeV} \; .
\end{equation}
With this choice of the Higgs mixing angles the states $\phi_j$ are
\CP mixtures.  Their reduced Yukawa couplings and reduced couplings to
$W,Z$ are given in \Tab{tab:coup1c}. The masses of $\phi_2$ and
$\phi_3$ are set to the values
\begin{align}\label{eq:mass-sc1c}
\text{scenario 1c:} & \quad m_2 = 400 \mbox{\,GeV}\;,\quad m_3 = 900
\mbox{\,GeV}\;.
\end{align}
The partial widths of the major decay modes of $\phi_2$ and $\phi_3$
and their total widths are listed in \Tab{tab:res_widths_sc1c}. Decay
modes whose width is smaller than a few $\times 10^{-3}$ GeV are not
listed. The given uncertainties result from the scale variations as
defined below \Eq{eq:rensc-var} in \Sec{subsec:setup}.
\begin{table}[htbp]
\centering
\caption{Values of the reduced couplings to fermions and  vector bosons $V=W,Z$
of the neutral Higgs bosons $\phi_j$ in scenario 1c.}\vspace{3mm}
{\renewcommand{\arraystretch}{1.2} \renewcommand{\tabcolsep}{0.2cm}
\begin{tabular}{lccccc}
  \hline \hline & $a_{jt}$ & $a_{jb}=a_{j\tau}$ & $b_{jt}$ &
  $b_{jb}=b_{j\tau}$ & $f_{jVV}$\\ \hline $\phi_1$ & 0.978 & 0.978 &
  0.208 & 0.208 & 0.978\\ $\phi_2$ & 0.560 & -0.854 & 0.692 & 0.692 &
  -0.147\\ $\phi_3$ & -0.854 & 0.560 & 0.692 & 0.692 & -0.147\\ \hline
  \hline
\end{tabular}}
\label{tab:coup1c}
\end{table}%
\vspace{2mm}
\begin{table}[htbp]
\begin{center}
\caption{Dominant partial decay widths and the total width of the \CP mixtures
$\phi_2$ and $\phi_3$ in scenario 1c.}  \vspace{3mm}
{\renewcommand{\arraystretch}{1.5} \renewcommand{\tabcolsep}{0.2cm}
\begin{tabular}{lll}
\hline\hline decay mode& $\Gamma_2$ [GeV] & $\Gamma_3$ [GeV]\\ \hline
 $\phi_i\rightarrow t\bar{t}$ & $8.47^{+0.18}_{-0.15}$ &
 $50.80^{-0.32}_{+0.27}$\\ $\phi_i\rightarrow VV$ & 0.52 & 7.37\\
 $\phi_i\rightarrow \phi_1Z$ & 0.27 & 4.73\\ $\phi_i\rightarrow
 \phi_2Z$ & 0 & 111.80\\ $\phi_i\rightarrow \phi_1\phi_1$ & 3.20 &
 6.14\\ $\phi_i\rightarrow \phi_1\phi_2$ & 0 & 4.00\\
 $\phi_i\rightarrow \phi_2\phi_2$ & 0 & 11.81\\ $\phi_i\rightarrow gg$
 & $0.030^{+0.005}_{-0.004}$ & $0.100^{+0.003}_{-0.005}$\\[1mm] \hline
 total & $12.49^{+0.19}_{-0.15}$ & $184.95^{-0.32}_{+0.26}$\\
 \hline\hline
\end{tabular}}
\label{tab:res_widths_sc1c}
\end{center}
\end{table}%

%

\subsubsection*{Scenario 2}

We choose again a \CP-conserving neutral Higgs sector scenario with
the same values of $m_1$, $m_+$, $\tan\beta$, and Higgs mixing angles
$\alpha_i$ as in scenarios 1a,b---cf.~\Eq{eq:angle-sc1ab}. Thus, the
reduced couplings of the $\phi_j$ given in \Tab{tab:Yuk1ab2} apply
also to this scenario.  The states $\phi_1$ and $\phi_2$ are \CP-even
while $\phi_3$ is \CP-odd. The masses of the two heavy neutral Higgs
bosons are set to the values
\begin{align}\label{eq:mass-sc2}
\text{scenario 2:} & \quad m_2 = 766 \mbox{\,GeV}\;,\quad m_3 = 750
\mbox{\,GeV}\;.
\end{align}
The total decay widths of the nearly mass-degenerate scalar and
pseudoscalar $\phi_2$ and $\phi_3$ are essentially determined by the
decay of these states to $\ttbar$, cf. \Tab{tab:res_widths_sc2}.
Again, decay modes whose width is smaller than a few $\times 10^{-3}$
GeV are not exhibited in this table.

\vspace{2mm}
\begin{table}[htbp]
  \begin{center}
\caption{Dominant partial decay widths and the total width of the \CP-even and
\CP-odd Higgs boson $\phi_2$ and $\phi_3$, respectively, in scenario
2.}  \vspace{3mm}
{\renewcommand{\arraystretch}{1.5} \renewcommand{\tabcolsep}{0.2cm}
\begin{tabular}{lll}
\hline\hline decay mode & $\Gamma_2$ [GeV] & $\Gamma_3$ [GeV]\\ \hline
$\phi_i\rightarrow t\bar{t}$ & $31.92^{-0.03}_{+0.02}$ &
$37.94^{-0.14}_{+0.12}$\\ $\phi_i\rightarrow gg$ &
$0.055^{+0.005}_{-0.005}$ & $0.087^{+0.006}_{-0.006}$\\[1mm] \hline
total & $31.97^{-0.02}_{+0.02}$ & $38.03^{-0.13}_{+0.11}$\\
\hline\hline
\end{tabular}}
\label{tab:res_widths_sc2}
\end{center}
\end{table}%

%

\subsubsection*{Experimental constraints}

The Yukawa couplings and couplings to the weak gauge bosons that are
assigned to the 125~GeV resonance in the above parameter scenarios are
in accord with the constraints from the LHC
\cite{Khachatryan:2016vau}. The strongest direct constraints on heavy
neutral Higgs bosons with strong couplings to top quarks were recently
reported by the ATLAS experiment \cite{ATLAS:2016pyq}. In this report
the 2HDM parameter region $\tan\beta<0.45$ $(\tan\beta<0.85)$ was
excluded at $95\%$ confidence level for a \CP-even (\CP-odd) Higgs
boson $H$ $(A)$ with a mass of 500 GeV. This analysis is based on
$\ttbar$ events that decay to leptons plus jets recorded at the LHC (8
TeV). The result supersedes previous bounds by ATLAS
\cite{Aad:2015fna} and CMS
\cite{Chatrchyan:2013lca,Khachatryan:2015sma}. Our parameter scenarios
1a,b,c above are not in direct conflict with the bounds of
\cite{ATLAS:2016pyq} because our choice $\tan\beta=1$ implies weaker
top-Yukawa couplings of $H$ and $A$ than those excluded in
\cite{ATLAS:2016pyq}: In the case of $H$ $(A)$ the squared top-Yukawa
coupling that enters the cross section for $gg\to \phi\to \ttbar$ is
reduced by a factor 0.20 (0.72). This suggests that a Higgs boson with
a mass of 400 GeV and top-Yukawa coupling strength as chosen in the
scenarios 1a,b,c is not yet excluded. We emphasize that the analysis
below does not crucially depend on the fact that we have chosen the
mass of the lighter of the two heavy states to be 400 GeV. The sole
purpose of these parameter choices is to illustrate with these
examples the sensitivity of top-spin observables to heavy Higgs
resonances.\\ The charged Higgs boson $H^\pm$ of the 2HDM plays no
decisive role in our analysis below. Our assumption $m_+ > 820$ GeV is
in accord with the non-observation of $H^\pm$ in direct searches at
the LHC and with the bounds derived from $B$ physics data
\cite{Mahmoudi:2009zx,Hermann:2012fc,Eberhardt:2013uba}.\\

The LHC data on the 125 GeV Higgs boson constrain CP-violating
top-Higgs couplings to some extent, see for instance
\cite{Kobakhidze:2016mfx}.  A direct search for $CP$ violation in
semileptonic $\ttbar$ events at the LHC(8 TeV) was recently reported
by the CMS experiment in \cite{Khachatryan:2016ngh}.  The set of spin
observables recently measured by the ATLAS experiment in dileptonic
$\ttbar$ events at 8 TeV \cite{Aaboud:2016bit} include $CP$-odd
observables and these measurements provide also direct bounds on $CP$
violation in $\ttbar$ production.  Our parameter scenario 1c is in
accord with the constraints from these analyses. Stronger, albeit
indirect constraints are obtained from low-energy data
\cite{Inoue:2014nva,Chen:2015gaa,Cirigliano:2016nyn,Cirigliano:2016njn},
in particular from the experimental upper limits on the electric
dipole moments (EDMs) of the neutron \cite{Baker:2006ts} and the
electron \cite{Baron:2013eja}. In the parameter scenario 1c the heavy
Higgs bosons have only a minor impact on these EDMs. The major
contribution results from $\phi_1$ exchange. The Yukawa and gauge
couplings of $\phi_1$ given in \Tab{tab:coup1c} lie within the allowed
parameter ranges derived in \cite{Chen:2015gaa,Cirigliano:2016njn}.

%

\section{Spin dependent observables}
\label{sec:obs}

In the following we study the impact of heavy Higgs bosons on the
(anti) top-quark polarization and top-antitop spin correlations,
taking NLO QCD corrections into account. We follow
\Ref{Bernreuther:2015yna} where a complete set of observables
sufficient to constrain the spin density matrix at the level of stable
top quarks has been presented. As in \Ref{Bernreuther:2015yna} we
analyze the angular distributions of the (anti-)top-quark decay
products.  More specifically, we consider dileptonic $\ttbar$ events
\begin{equation} \label{eq:ttdilept}
  p p \longrightarrow t + {\bar t} + X \longrightarrow \ell^+~\ell'^-
  + \mbox{jets} + E_T^{\mbox{\scriptsize miss}} \, , \quad \ell,~\ell'=e,\mu ,
\end{equation}
and define four dileptonic angular correlations that correspond, at
the level of the intermediate $t$ and $\bar t$ quarks, to $P$- and
\CP-even \ttbar spin correlations. In addition, we consider for the
reactions \eqref{eq:ttdilept} also a $P$-and \CP-odd triple
correlation and a lepton angular distribution that corresponds to the
longitudinal top-quark polarization. Moreover, we analyze the
correlation of the azimuthal angles of the charged leptons and
investigate its sensitivity to the \CP~nature of the heavy Higgs
bosons.

%

\subsection{Observables sensitive to top-antitop spin correlations}
\label{subsec:spincorr}
We consider for (\ref{eq:ttdilept}) the following normalized differential distribution of the outgoing leptons:
\begin{equation}
  \label{eq:DoubleDiffDistribution}
  \frac{1}{ \sigma} \frac{d^2\sigma}{
    d\cos\theta_{+}\,d\cos\theta_{-}} = \frac{1}{4}\left(1+ B_1
    \cos(\theta_{+}) + B_2 \cos(\theta_{-}) -
    C\cos(\theta_{+})\cos(\theta_{-}) \right),
\end{equation}
where $\theta_+$ $(\theta_-)$ denotes the angle between the direction
of flight $\hl_+$ ($\hl_-$) of the positively (negatively) charged
lepton in the (anti-)top-quark rest frame\footnote{The respective rest
frames are reached through a rotation-free boost from the $\ttbar$
zero-momentum frame.}  and a reference axis $\ha$ ($\hb$) to be
defined below ($\theta_+=\angle(\hl_+,\ha)$,
$\theta_-=\angle(\hl_-,\hb)$).  We assume that the top-quarks decay
dominantly via the SM decay $t\to W^+b$ ($\bar t \to W^-\bar b$) with
W-boson decaying further into a lepton neutrino pair.  The coefficients
$B_{1}$, $B_{2}$ and $C$ depend on the chosen reference axes $\ha$ and
$\hb$. If no acceptance cuts are applied, $C$ is related to the double
spin asymmetry at the level of the intermediate top quarks:
\begin{equation}
  \label{eq:spinasymmetry}
  C_{\text{ab}} \equiv C(\ha,\hb) =
  \kappa_\ell^2\,\frac{\sigma_{\ttbar}\,(\uparrow\uparrow)
    +\sigma_{\ttbar}\,(\downarrow\downarrow)
  -\sigma_{\ttbar}\,(\uparrow\downarrow)
  -\sigma_{\ttbar}\,(\downarrow\uparrow)}{
  \sigma_{\ttbar}\,(\uparrow\uparrow)
  +\sigma_{\ttbar}\,(\downarrow\downarrow)
  +\sigma_{\ttbar}\,(\uparrow\downarrow)
  +\sigma_{\ttbar}\,(\downarrow\uparrow)}.
\end{equation}
$\sigma_{\ttbar}$ denotes the cross section for top-quark pair production.
The first (second) arrow refers to the spin state of the $t$ ($\bar
t$) quark with respect to the axis $\ha$ ($\hb$). The prefactor
$\kappa_\ell^2$ is due to the spin analyzer quality of the decay
products. In the conventions used here
\begin{displaymath}
  \kappa_\ell = \kappa_{{\ell^+}} = \kappa_{{\ell^-}}.
\end{displaymath}
At NLO QCD its value is $\kappa_\ell=1-0.015\alpha_s$
\cite{Czarnecki:1990pe}.  Following
\Ref{Bernreuther:2015yna} we use the beam direction $\hp =(0,0,1)$ and
the direction of flight of the top-quark $\hk$ in the $t\bar t$
zero-momentum frame (ZMF) to construct an orthonormal basis:
\begin{equation}
  \{\hk,\hn,\hr\}:\quad \hr = \frac{1}{\R} (\hp -\Y \hk)\,, \quad \hn
  =\frac{1}{\R} (\hp \times \hk), \label{proONB}
\end{equation}
with
\begin{displaymath}
  \Y = \hp\cdot \hk\,, \quad \R=\sqrt{1-\Y^2}\, .
\end{displaymath}

Using the orthonormal basis $\{\hk,\hn,\hr\}$ the reference axes $\ha$
and $\hb$ are defined in \Tab{tab:axes}. The factor $\sign(y)$ takes
the Bose symmetry of the initial $gg$ state into account.
%
\begin{table}[htbp]
  \caption{Choice of reference axes. The unit vectors $\hn$, $\hr$ and the
    variable $\Y$ are defined in \Eq{proONB}. The factors
    $\text{sign}(\Y)$ are required because of the Bose symmetry of the
    initial $gg$ state.  \vspace{0.1cm}}
  \label{tab:axes}
  \centering {\renewcommand{\arraystretch}{1.3}
  \renewcommand{\tabcolsep}{0.2cm}
    \begin{tabular}{cccc}
      \hline \hline & Label & $\ha$ & $\hb$\\ \hline transverse & n &
      sign$(\Y)\; \hn$ & $-$sign$(\Y)\; \hn$\\ r axis & r &
      sign$(\Y)\; \hr$ & $-$sign$(\Y)\; \hr$\\ helicity & k & $\hk$ &
      $-\hk$\\ \hline \hline
    \end{tabular}}
\end{table}%
Using the different reference axes as shown in \Tab{tab:axes} nine
different correlations $C(\ha,\hb)$ can be defined. 
As far as $P$- and $CP$-even correlations are concerned   we restrict
ourselves to the diagonal correlations
\begin{displaymath}
  \Ckk,\mbox{ } \Cnn,\mbox{ and } \Crr \, .
\end{displaymath}
 These correlations are sensitive to the $P$- and \CP-even
contributions of the spin density matrix, do not require absorptive
parts, and receive Higgs-boson contributions at leading order.
Since the choice $\ha=\hk$ ($\hb=-\hk$) is equivalent to the
quantization axis in the helicity basis, $\Ckk$ is often abbreviated
in the literature as $\Chel$. In the following we adopt this notation
and use $\Chel$ instead of $\Ckk$.  Instead of measuring the double
differential distribution as given in \Eq{eq:DoubleDiffDistribution}
the coefficient $C$ can also be determined from the expectation value of
$\cos(\theta_+) \cos(\theta_-)$:
\begin{equation}
  \langle \cos(\theta_+) \cos(\theta_-)\rangle = -\frac{1}{9} C.
\end{equation}
As pointed out in \Ref{Bernreuther:2015yna} the correlations $\Chel,
\Cnn,$ and $\Crr$ are related to the opening angle distribution
\begin{equation}
  \frac{1}{\sigma}\frac{d\sigma}{d\cos\varphi}=\frac{1}{2}
\left(1-D\cos\varphi\right),
\end{equation}
where the angle $\varphi$ is defined by $\varphi =
\angle(\hl_+,\hl_-)$.  Since the vectors $\{\hk,\hn,\hr\}$ form an
orthonormal basis the relation
\begin{equation}
  D=-\frac{1}{3}(\Chel+\Cnn+\Crr)
  \label{D1}
\end{equation}
holds. \Eq{D1} may be used to cross check results.  $D$ can also be
determined from the expectation value of $\cos\varphi$:
\begin{equation}
  D = -3 \langle \cos(\varphi)\rangle.
\end{equation}
For completeness we note that both $C_{\text{ab}}$ and $D$ are related
to expectation values of spin observables at the level of the
intermediate top quarks:
\begin{equation}
  C_{\text{ab}} = \kappa_\ell^2 4\langle(\St\cdot \ha) 
  (\Stbar\cdot \hb)\rangle, 
\end{equation}
\begin{equation}
  D = \kappa_\ell^2 \frac{4}{3}\langle(\St\cdot\Stbar)\rangle, 
\end{equation}
where $\St$ ($\Stbar$) denotes the (anti-)top-quark spin
operator.
%
%

\subsection{Observables sensitive to top-quark and antiquark polarization}
\label{subsec:pol}

Integrating out the angle of the positively/negatively charged lepton
in \Eq{eq:DoubleDiffDistribution} leads to a single differential
distribution:
\begin{equation}
  \label{eq:SingleDiffDistribution}
  \frac{1}{ \sigma} \frac{d\sigma}{ d\cos\theta_{\pm}} =
    \frac{1}{2}\left(1+ B_{1,2} \cos(\theta_{\pm}) \right).
\end{equation}
The coefficient $B_{1}$ ($B_2$) is related to the polarization of the
intermediate (anti-)top-quark:
\begin{equation}
  B_1(\ha) = \kappa_\ell \langle 2\St\cdot \ha\rangle, \quad B_2(\hb)
  = -\kappa_\ell \langle 2\Stbar\cdot \hb\rangle.
\end{equation}
 (The polarization is defined as twice the expectation value
of the spin operator: $\mathbf{P}_t = 2\langle\St\rangle$.)  For $\ha
= \hk$ ($\hb=-\hk$) the coefficient $B_1$ ($B_2$) is a measure for the
longitudinal polarization of the intermediate (anti-)top-quark. Note
that a non-vanishing 
\begin{equation}
  B_t = B_{1}(\hk),\quad B_{\bar t} = B_{2}(-\hk)
\end{equation}
requires a parity violating
interaction. As a consequence, within the SM only the parity-violating
weak interactions generate a small non-zero coefficient $B_1(\hk)$ of less
than 1 \%.  Similar to $D$ the coefficients $B_{1,2}$ are related to
expectation values of the respective angles:
\begin{equation}
B_{1,2}=3\langle\cos\theta_\pm\rangle.
\end{equation}
We note that the single differential distributions can also be studied
for top-quark pairs decaying semi-leptonically. If no acceptance cuts
are applied, the coefficient $B_1$ ($B_2$) parametrizing the
distribution of the positively (negatively) charged lepton is the same
in both channels.
%

\subsection{Triple correlation}
\label{subsec:trip}
In addition to the aforementioned observables we analyze the $P$- and
\CP-odd triple correlation:
\begin{equation}
  \label{eq:OCP} \OCP = (\hlp\times\hlm)\cdot\hk.
\end{equation}
The observable $\OCP$ is sensitive to $P$- and $CP$-odd (dispersive)
new physics contributions. Within the scenarios considered here only
scenario 1c leads to a non-zero expectation value.  As discussed in
\Ref{Bernreuther:2015yna} $\OCP$ is related to a linear combination of
$\Cnr$ and $\Crn$:
\begin{equation}
  \label{eq:OCPvsC}
  \Cnr-\Crn = 9 \langle \OCP \rangle.
\end{equation}
Up to prefactors it can also be related to a spin observable at the
level of the intermediate top-quarks:
\begin{equation}
  \label{eq:OCPvsSpinObs}
  \langle\OCP\rangle = -\frac{\kappa_\ell^2}{9}\langle
  \left(\St\times\Stx\right)\cdot\hk\rangle\,.
\end{equation}
One may consider also other \CP-odd dileptonic triple correlations,
for instance, a correlation where $\hk$ in \Eq{eq:OCP} is replaced
by the proton beam direction, i.e. $\mbox{sign}(\Y)\hp$. 
But this correlation has a
very low sensitivity to resonant Higgs-boson induced \CP violation,
as an inspection of the corresponding squared $S$ matrix element
shows.

%
%

\subsection{Difference between the leptonic azimuthal angles}
\label{subsec:dphi}

Following \Ref{Berge:2014sra} we construct a \CP-sensitive angular
observable $\dphi$ as follows. We identify the $z$-axis in the top and
antitop rest frames with the direction of flight of the top quark in
the \ttbar ZMF, $\hk$. With respect to this $z$-axis we define the
azimuthal angle $\phi^*$ between the charged leptons
\begin{equation}
\phi^*=\arccos(\hlp^{\perp}\cdot\hlm^{\perp})\,,\quad\phi^*\in[0,\pi]\,.
\end{equation}
The unit 3-vector $\hlp^{\perp}$ ($\hlm^{\perp}$) defines the
direction of the antilepton (lepton) perpendicular to $\hk$ in the top
(antitop) rest frame ($\hlpm^{\perp}\cdot\hk=0$):
\begin{equation}
\hlpm^{\perp}
=\frac{\hlpm-(\hlpm\cdot\hk)\hk}{\left|\hlpm-(\hlpm\cdot\hk)\hk\right|}\,.
\end{equation}
Similar to \Ref{Berge:2014sra} we use the \CP-odd triple product,
$(\hlp\times\hlm)\cdot\hk$, to construct a \CP-sensitive observable
from $\phi^*$ in order to probe the \CP properties of the heavy Higgs
bosons:
\begin{equation}
\dphi=\left\{\begin{array}{l@{\quad\text{if}\quad}r} \phi^* &
(\hlp\times\hlm)\cdot\hk\ge0\\ 2\pi-\phi^* &
(\hlp\times\hlm)\cdot\hk<0
\end{array}\right.\,,\quad\dphi\in[0,2\pi]\,.
\label{dphi}
\end{equation}
We are interested in the (normalized) distribution of this observable
$\sigma^{-1}d\sigma/d\dphi$ because of its potential to discriminate
between \CP-even, \CP-odd and \CP-mixed heavy Higgs bosons. The
parameters of scenarios 1a--1c (cf. \Sec{sec:2hdm}) are chosen such
that these cases can be directly compared and the sensitivity of
$\dphi$ to the \CP-properties of the heavy Higgs coupling to top
quarks can be studied in detail.

%

\section{Results}
\label{sec:res}

In this section we briefly describe the setup of our calculation.
Furthermore, we present phenomenological results at NLO QCD for the
distribution of the cross section with respect to the invariant mass
of the top-quark pair and for the spin dependent observables
introduced in the previous section.

%

\subsection{Setup}
\label{subsec:setup}
For all observables introduced in the previous section we make
predictions within the SM including QCD and weak corrections as well
as predictions including the two additional (neutral) heavy Higgs
bosons of the 2HDM parameter scenarios of Sec.~\ref{sec:2hdm}.  We
stress that we take into account also the interference of the signal
and background contributions at NLO. The contribution of the 125 GeV
Higgs resonance is included in the SM predictions.

As far as the NLO QCD corrections are concerned we apply the same
approximations to the non-SM contributions as we did in
\Ref{Bernreuther:2015fts}. We use the heavy top mass limit including
an effective K-factor rescaling  to
obtain the leading resonant contributions at NLO. For details we refer
to \Ref{Bernreuther:2015fts}.

The SM input parameters are chosen as in
\Ref{Bernreuther:2015fts}. The top-quark mass renormalized in the
on-shell scheme is set to
\begin{equation}
  m_t = 173.34\text{ GeV}.
\end{equation}
For the values of the electromagnetic fine structure constant $\alpha$
and the gauge boson masses we use
\begin{equation}
  \alpha = \frac{1}{129},\quad m_W = 80.385\text{ GeV}, \quad m_Z =
  91.1876\text{ GeV}.
\end{equation}
For the parton distribution functions (PDFs) we employ the PDF set
CT10nlo \cite{Lai:2010vv} which provides also the value of the strong
coupling $\alpha_s(\mur)$ at the renormalization scale $\mur$.  As
central scale we chose
\begin{equation}
  \mur = \muf = \mu_0
\end{equation}
($\muf$ denotes the factorization scale) with
\begin{equation}
  \label{eq:rensc-var}
  \mu_0=\frac{m_2+m_3}{4}\,,
\end{equation}
motivated by the choice $\mu_0=m_H/2$ in the SM case (see, for
instance, \cite{Dittmaier:2011ti}).  The uncertainties due to the
residual dependence of the theoretical predictions on $\muf$ and
$\mur$ are estimated by varying both scales simultaneously
($\mur=\muf\equiv\mu$) by a factor of two up and down.  All results in
this section are presented for proton-proton collisions at a center of
mass energy of $\sqrt{s}=13$ TeV and are thus applicable to the data
collected during the LHC run II.

When calculating theoretical predictions for the observables defined
in the previous section we always expand all ratios or normalized
cross sections in the coupling constants. This is a consistent
approximation when working in fixed order perturbation theory.
%

\subsection{Top-quark pair invariant mass distribution}
\label{subsec:mttbar}

The invariant mass distribution of the $\ttbar$ pair,
$d\sigma_{\ttbar}/dM_{\ttbar}$ 
where  $\mtt=\sqrt{(k_t+k_{\bar t})^2}$,
is the basic observable in the search for heavy (Higgs) resonances 
in the $\ttbar$ channel. Here we compute this distribution at NLO in
the SM 
and in the 2HDM using the parameter scenarios of Sec.~\ref{sec:2hdm}.
(These parameter sets are different from those used in
\Ref{Bernreuther:2015fts}.) In \Sec{subsec:spin} we analyze top-spin
dependent observables and compare the sensitivities of these
observables and $d\sigma_{\ttbar}/dM_{\ttbar}$ to heavy Higgs boson effects.

\begin{figure}[htbp]
  \begin{center}
  \includegraphics[scale=1]{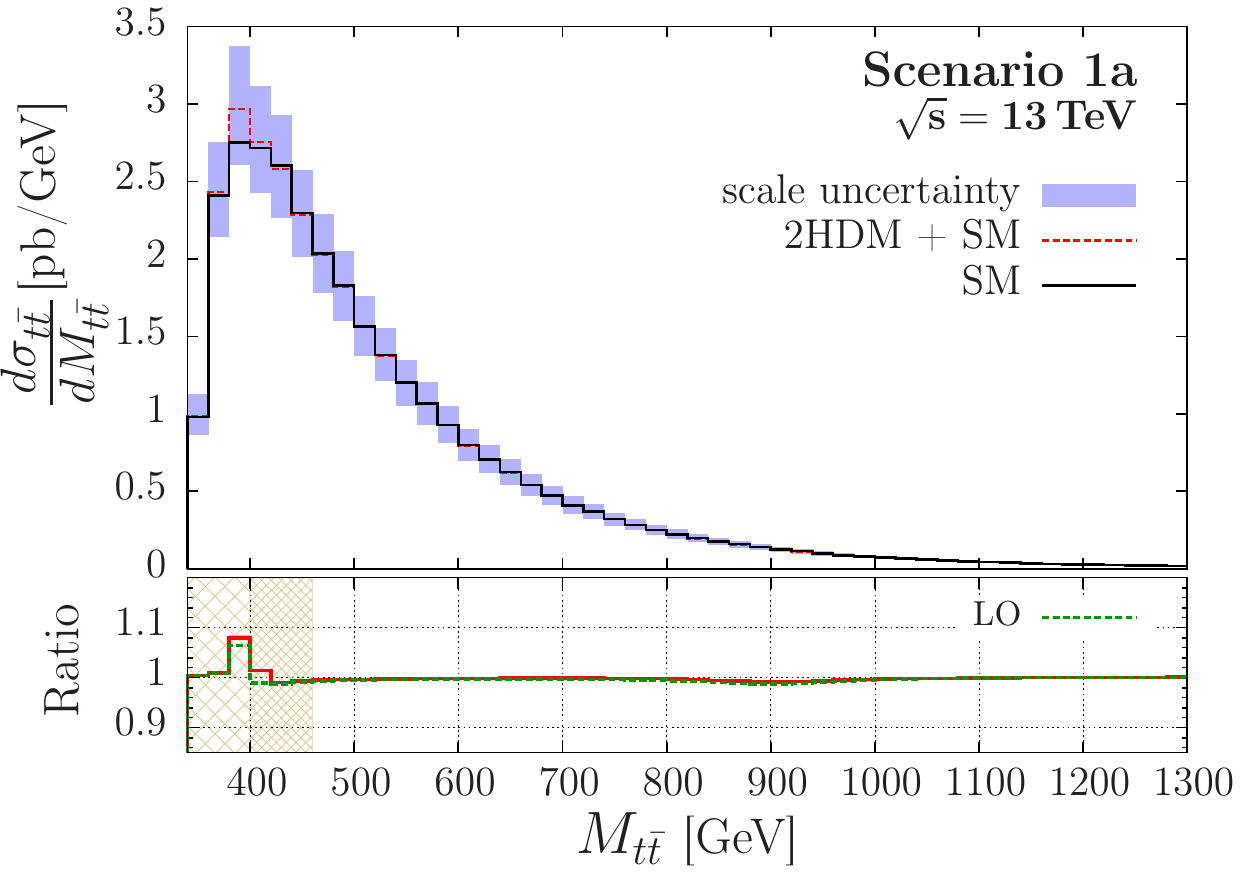}
  \end{center}
  \caption{Distribution of the $\ttbar$ invariant mass, \mtt,  at NLO 
    for scenario 1a. The upper pane shows the SM contribution (solid
    black) including NLO QCD and weak corrections and the sum of SM
    and 2HDM contributions (dashed red) at NLO QCD.  The blue shaded
    area represents the scale uncertainty. The lower pane show the
    ratios of the \mtt distribution for the SM + 2HDM and SM at LO
    (dashed green) and NLO (solid red). The hatched regions in the
    ratio plots display the \mtt bins which are used to evaluate the
    spin dependent observables; cf. Sec.~\ref{subsec:spin}.  }
  \label{fig:mttbar1a}
\end{figure}%
\begin{figure}[htbp]
  \begin{center}
  \includegraphics[scale=1]{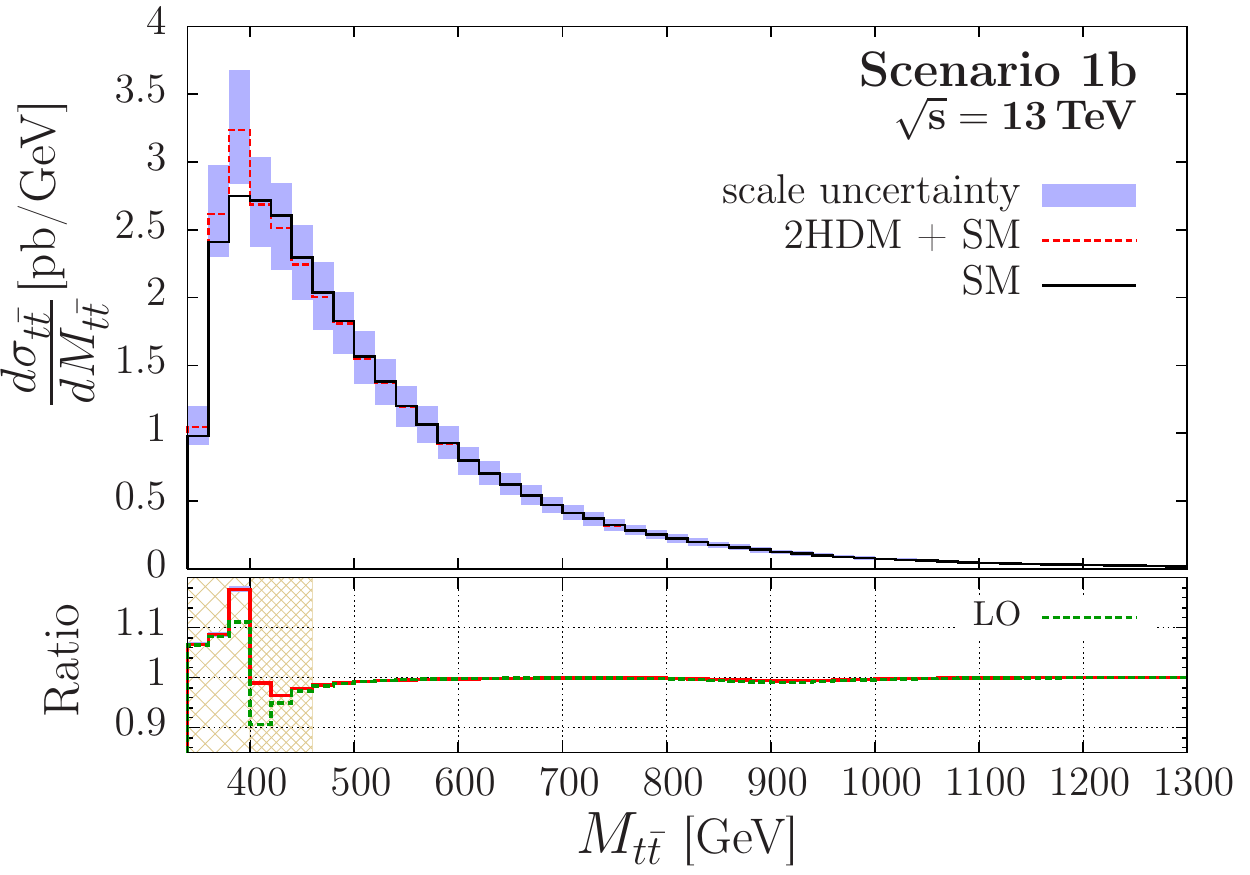}%
  \end{center}
  \caption{Same as \Fig{fig:mttbar1a} but for scenario 1b.}
  \label{fig:mttbar1b}
\end{figure}%
\begin{figure}[htbp]
  \begin{center}
  \includegraphics[scale=1]{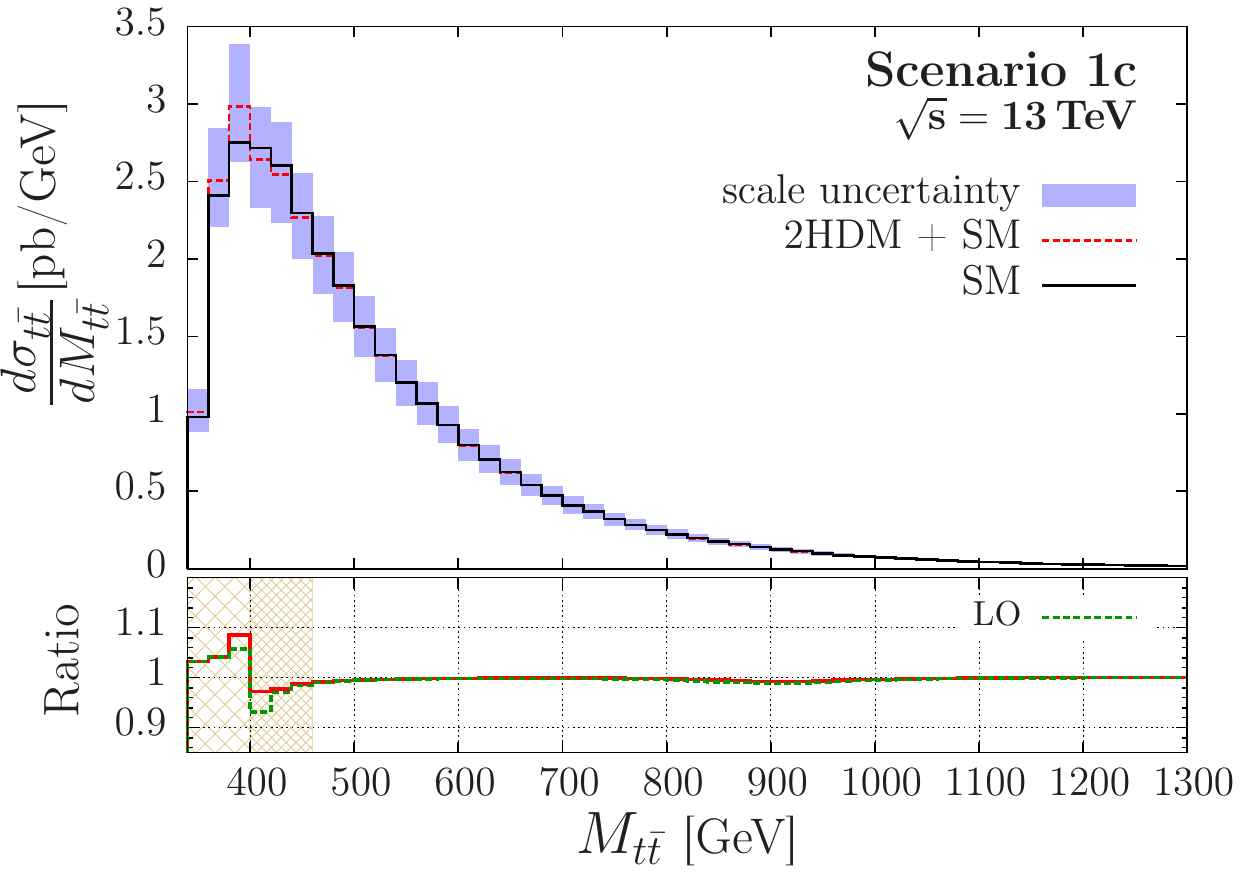}%
  \end{center}
  \caption{Same as \Fig{fig:mttbar1a} but for scenario 1c.}
  \label{fig:mttbar1c}
\end{figure}%
\begin{figure}[htbp]
  \begin{center}
  \includegraphics[scale=1]{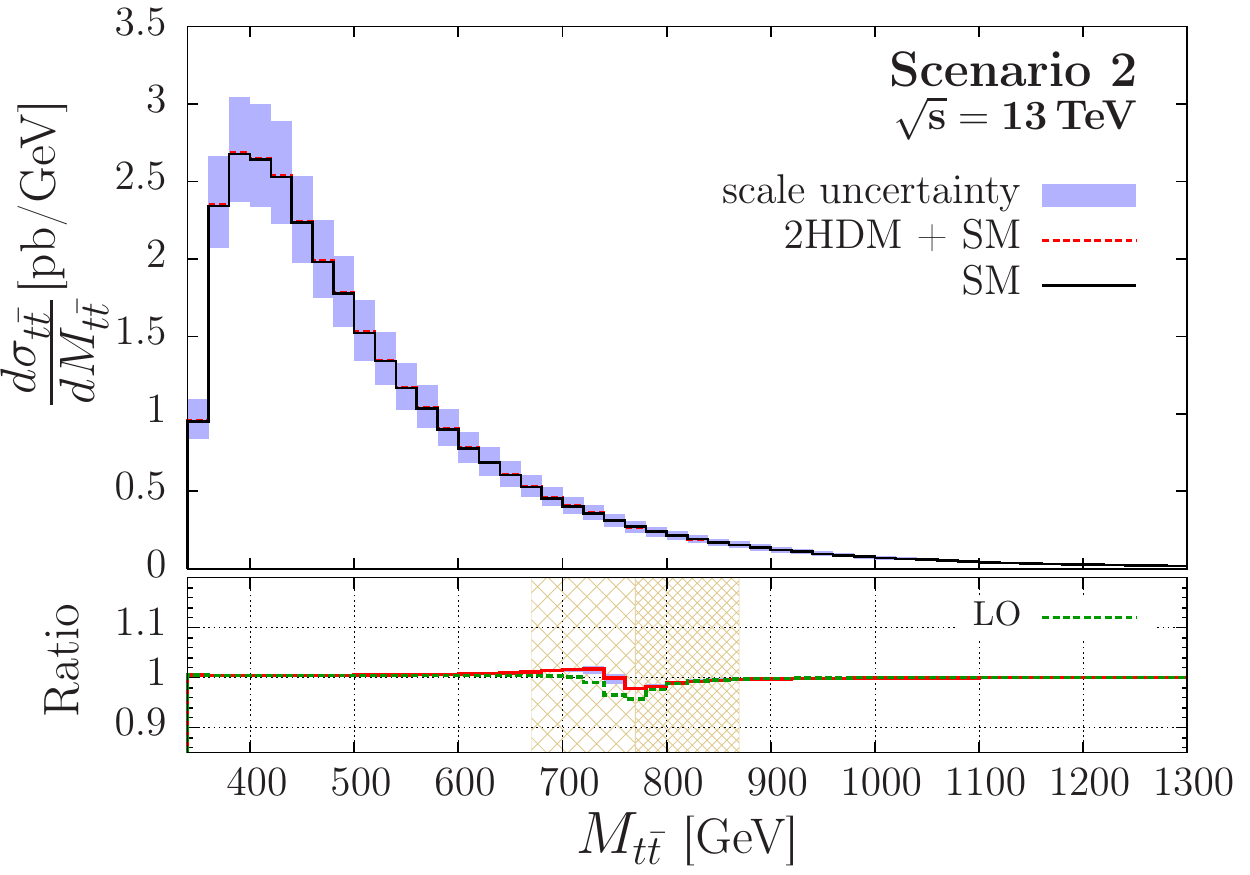}%
  \end{center}
  \caption{Same as \Fig{fig:mttbar1a} but for scenario 2.}
  \label{fig:mttbar2}
\end{figure}%

The top-quark pair invariant mass distributions are displayed for
scenarios 1a--1c and 2 in \Figs{fig:mttbar1a}--\ref{fig:mttbar2}.  For
scenarios 1a--1c the peak-dip structure that results from the
interference between signal and background is clearly visible for the
lighter resonance at 400 GeV. For scenario 1b where the lighter
resonance is a pseudoscalar we even observe a signal-to-background
(S/B) ratio of about 18\% in the \mtt bin from 380 GeV to 400
GeV. However, a bin width as small as 20 GeV cannot be attained by
present experiments.  For experimentally achievable bin widths of
about 40~GeV (see for example \Ref{ATLAS:2016pyq}) in the lower \mtt
range, the effect is 8\%--13\% depending on the bin position.  In view
of the recent ATLAS analysis \cite{ATLAS:2016pyq} scenario 1b may be
excluded. However, our main interest in this scenario is mostly to
compare top spin effects induced by different heavy Higgs boson \CP
eigenstates and \CP-mixed states and study the potential gain in
sensitivity with respect to top-spin independent observables. The
findings concerning the sensitivity of top-spin observables may also
apply to other parameter scenarios.  The resonance at 900 GeV in
scenarios 1a--1c is barely visible in the $\mtt$-distribution, mainly
because of its large decay width.\\ In scenario 2 the heavy scalar and
pseudoscalar resonances overlap and generate a single resonance
structure between 700 GeV and 800 GeV. It is much more pronounced than
the 900 GeV resonance in scenarios 1a--1c because of the smaller decay
widths of the heavy Higgs bosons in scenario 2.  The parameters of
scenario 2 were chosen to allow a direct comparison with
\Ref{Djouadi:2016ack} where higher-order QCD corrections to the signal
and interference have been estimated by applying a constant K-factor,
$\text{K}=2$.  As our analysis shows, a simple K-factor rescaling as
applied in \Ref{Djouadi:2016ack} is questionable and is in general not
sufficient to account for higher order corrections to the $\mtt$
distribution. In fact, within the approximations of the NLO
corrections used here, the dip between 700~GeV and 800~GeV observed at
LO (dashed-green line in \Fig{fig:mttbar2}) 
is reduced at NLO (solid-red line in \Fig{fig:mttbar2}). 
Hence, an experimental analysis using the naive
K-factor rescaling would in this case lead to experimental bounds that
are too strong---provided no effect will be seen.\\
We choose in the next section, as in \Ref{Bernreuther:2015fts},
appropriate \mtt bins for evaluating the top-spin dependent
observables, in order to avoid a cancellation due to the peak-dip
structure in the \mtt spectrum.  Thus we estimate, for the chosen
parameter scenarios, with an optimized binning the maximal effects of
heavy Higgs boson resonances in the \ttbar decay channel.  In
\Figs{fig:mttbar1a}--\ref{fig:mttbar2} these \mtt bins are indicated
by the hatched regions in the ratio plots.  Since in an actual
experimental analysis the optimal bin locations are unknown---unless a
significant distortion in the \mtt distribution will be found---one
would study the observables as functions of the bin location.

\subsection{Spin dependent observables}
\label{subsec:spin}
\begin{figure}[htbp]
  \begin{center}
    \includegraphics[scale=1]{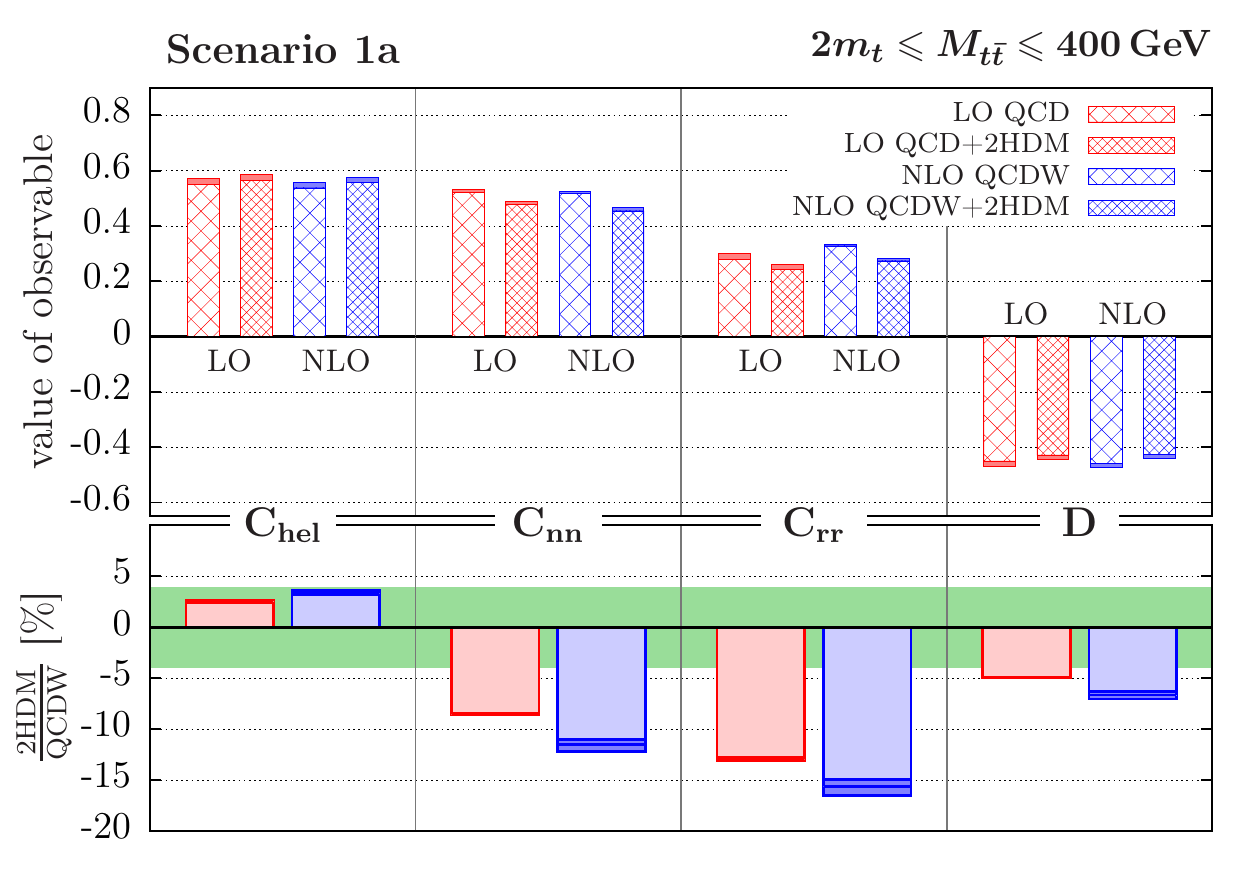}\\
    \includegraphics[scale=1]{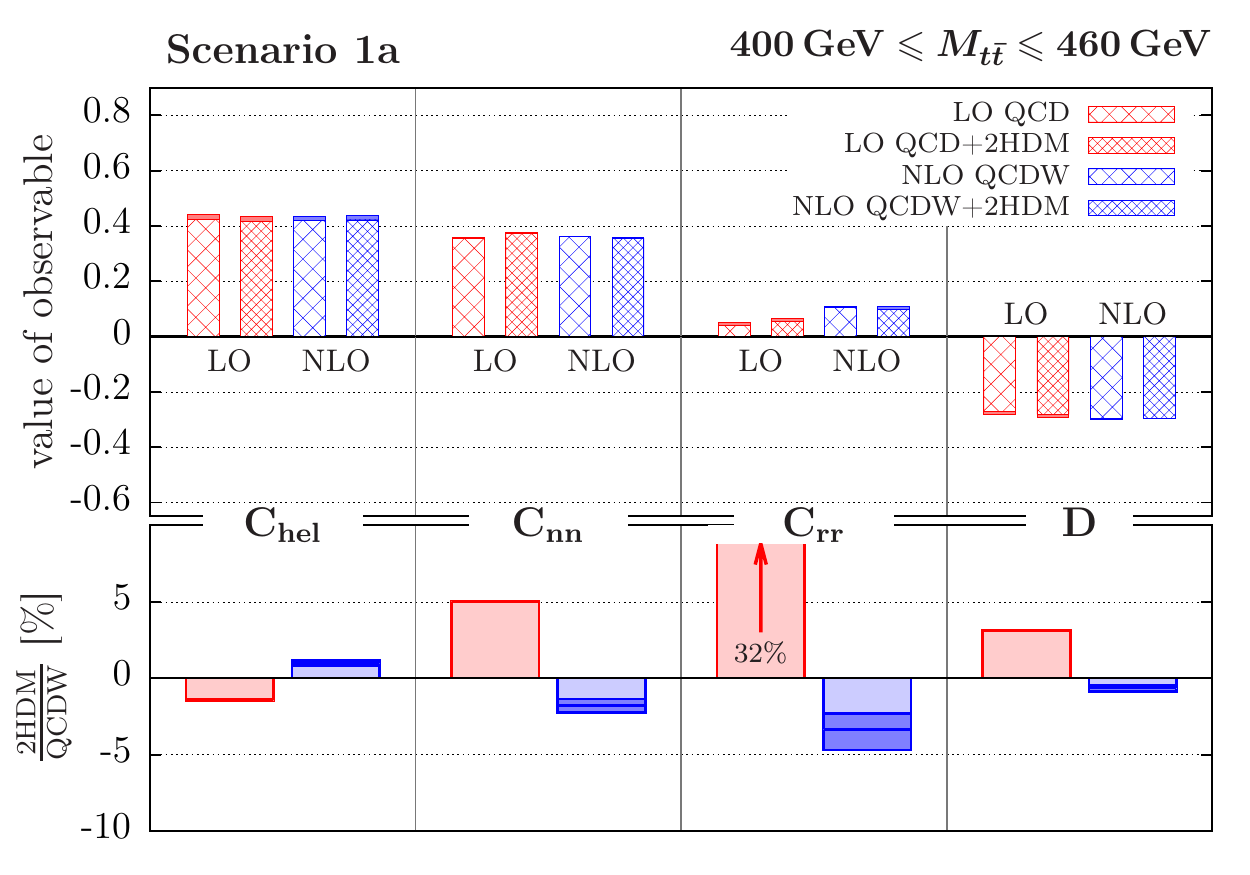}
    \end{center}
\caption{$P$- and $CP$-even spin correlations in scenario 1a for the LHC (13 TeV).}
\label{fig:spincorr_1a}
\end{figure}%
\begin{figure}[htbp]
  \begin{center}
    \includegraphics[scale=1]{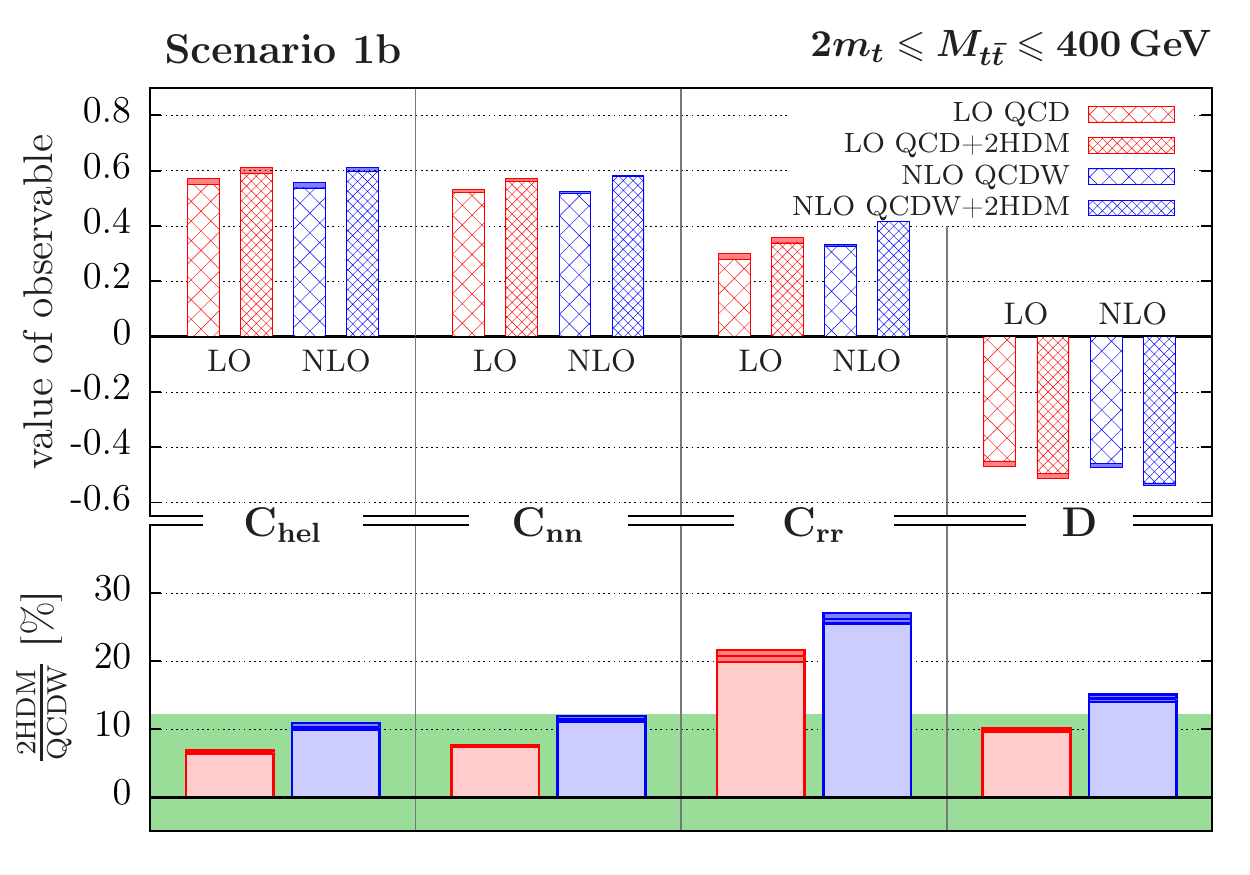}
\includegraphics[scale=1]{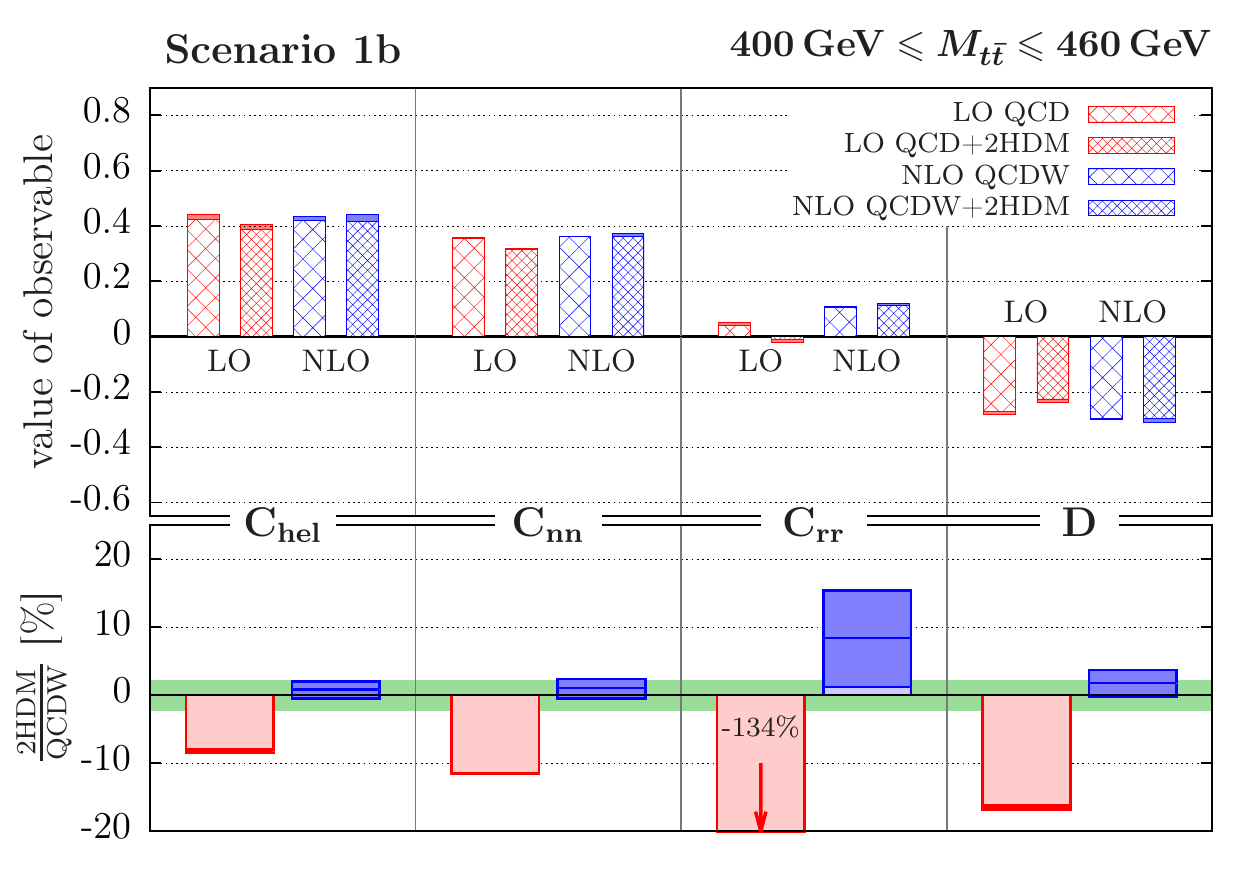}
  \end{center}
\caption{ $P$- and $CP$-even spin correlations in scenario 1b for the LHC (13 TeV).}
\label{fig:spincorr_1b}
\end{figure}%
\begin{figure}[htbp]
  \begin{center}
    \includegraphics[scale=1.]{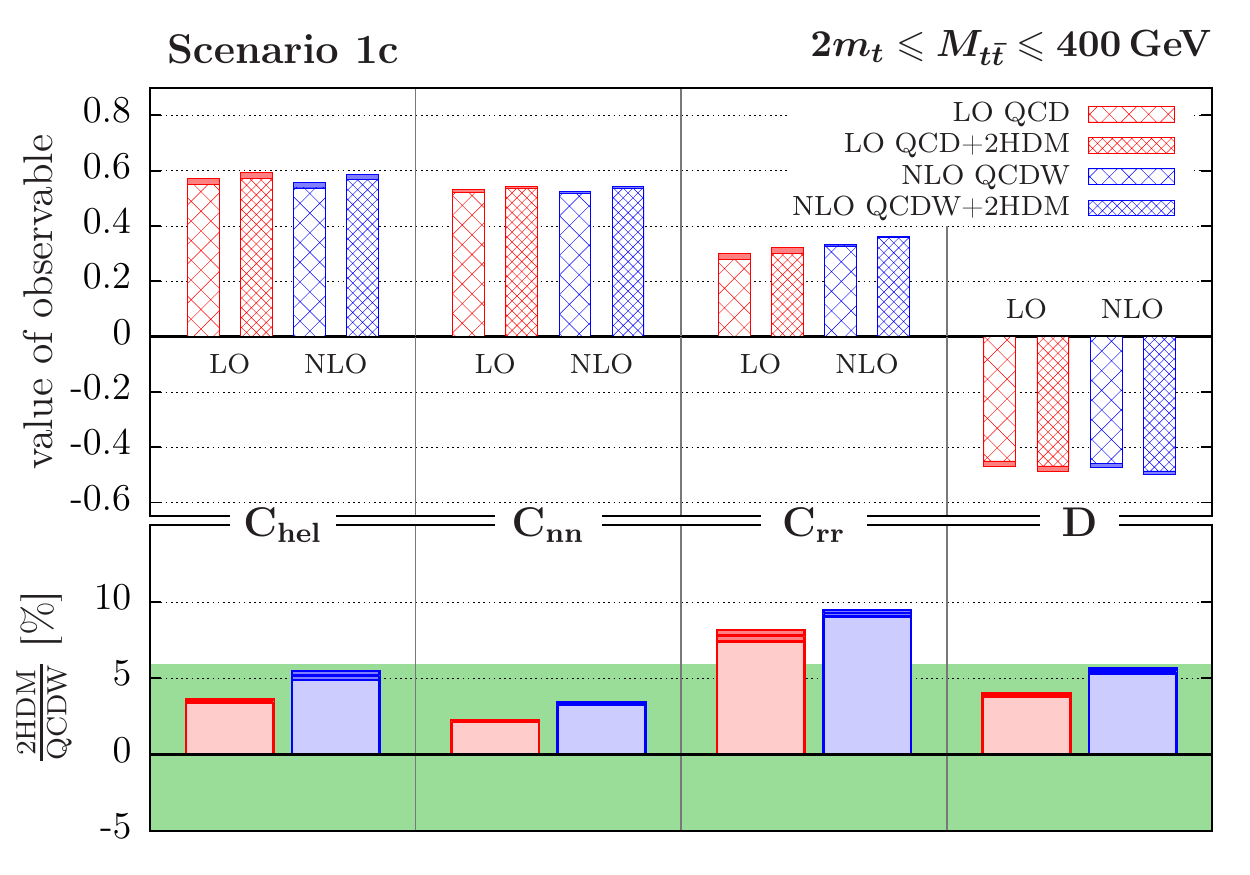}
    \includegraphics[scale=1.]{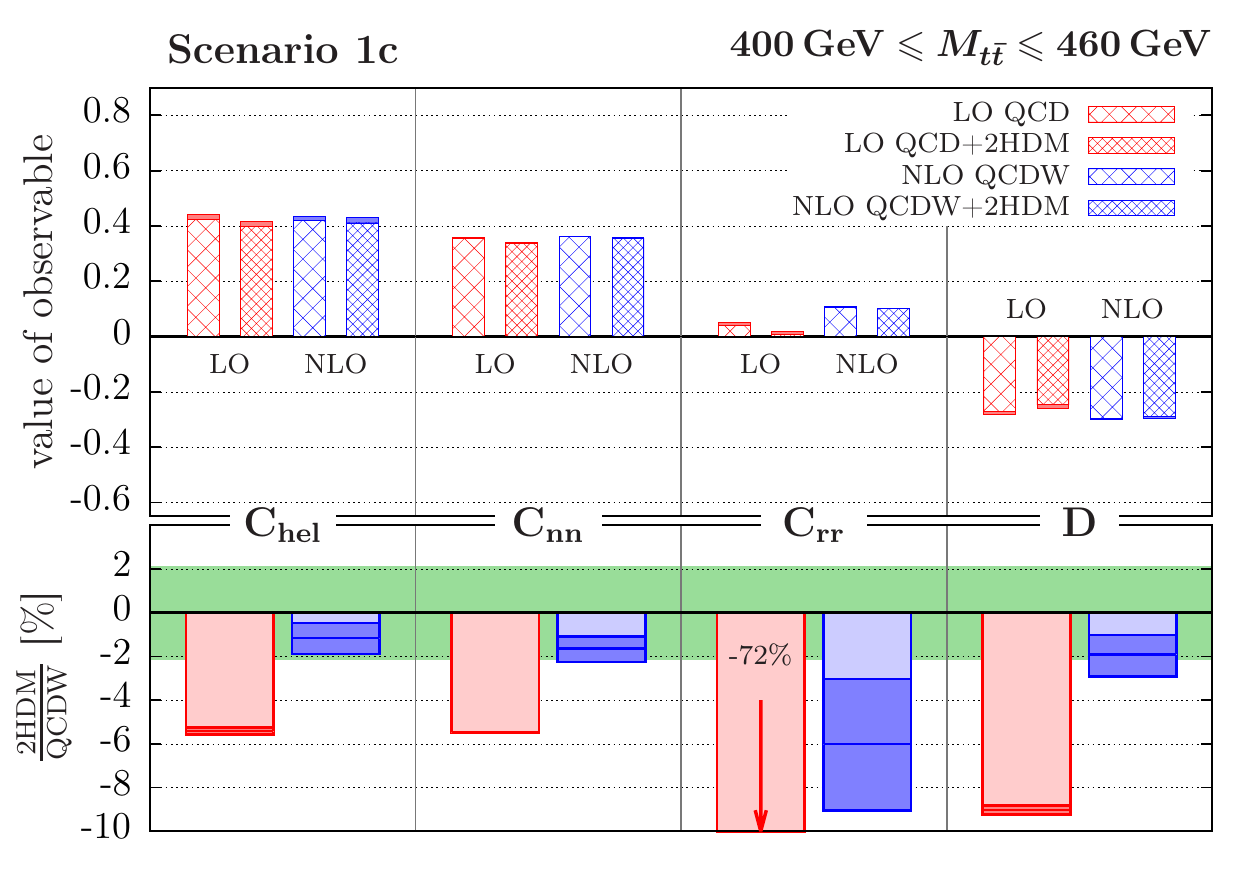}
  \end{center}
\caption{$P$- and $CP$-even spin correlations in scenario 1c for the LHC (13 TeV).}
\label{fig:spincorr_1c}
\end{figure}%

\begin{figure}
  \begin{center}
    \includegraphics[scale=1]{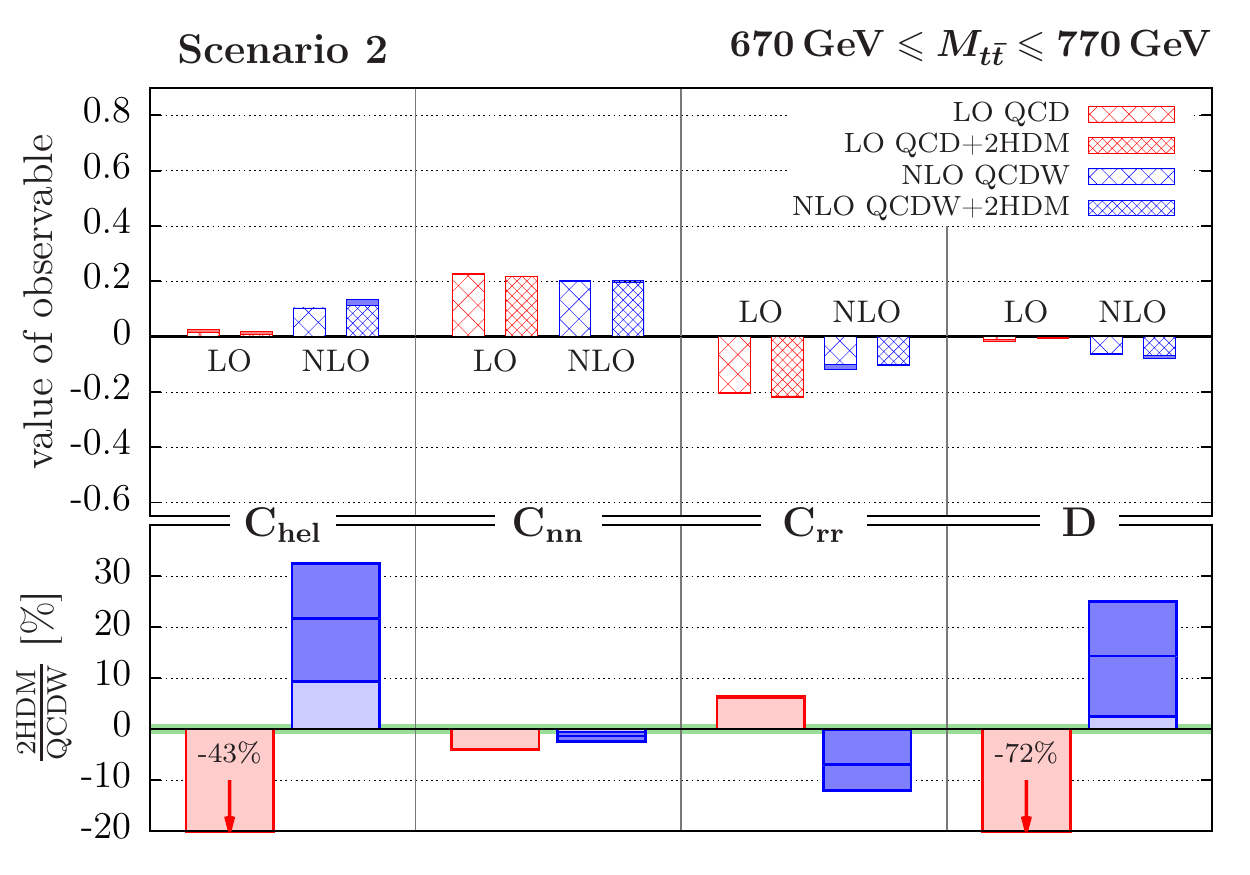}
    \includegraphics[scale=1]{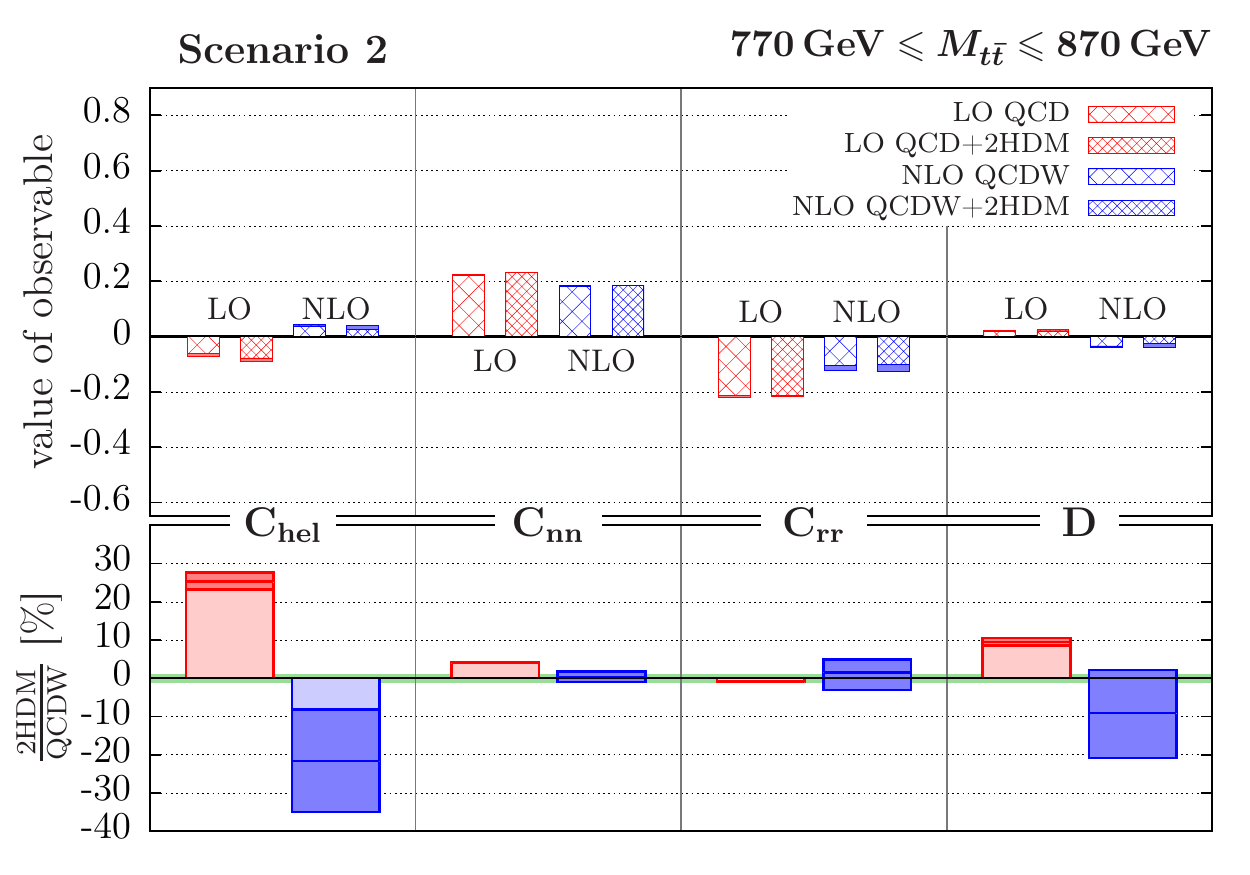}
  \end{center}
\caption{$P$- and $CP$-even spin correlations in scenario 2 for the LHC (13 TeV).}
\label{fig:spincorr_2}
\end{figure}

We consider now the top-spin dependent observables introduced in
 Sec.~\ref{sec:obs} for the dileptonic \ttbar events
 \eqref{eq:ttdilept}.  As discussed in \Sec{subsec:mttbar}, we
 evaluate these observables within two \mtt bins in order to enhance
 the S/B ratio, namely $2m_t$--400 GeV and 400--460 GeV in scenarios
 1a--1c and 670--770 GeV and 770-870 GeV in scenario 2. Results
 for the lepton angular correlations \Chel, \Cnn, \Crr and \D that
 correspond to $P$- and $CP$-even \ttbar spin correlations are shown
 in Figs.~\ref{fig:spincorr_1a}--\ref{fig:spincorr_1c} for scenarios
 1a--1c, respectively, and in Fig.~\ref{fig:spincorr_2} for scenario
 2. In the upper panel of each plot the values of \Chel, \Cnn, \Crr
 and \D are displayed for the SM (coarse hatched) and for the SM +
 2HDM contributions (fine hatched) including interference  at LO (red)
 and NLO (blue). The solid filled regions of the bars represent the
 scale uncertainty estimated by varying the scale as described in
 \Sec{subsec:setup}.

In the lower panel of each plot the S/B ratio of the 2HDM contribution
(including interference) and the SM prediction is shown at LO (red)
and at NLO (blue).  The darker parts of the bars represent the scale
uncertainty of this ratio. In cases where the ratio takes on extreme
values, the bars are only shown partially and the central value of the
respective ratio is then given in the plot.  The green shaded area in
the ratio plots display, for comparison, the S/B ratio of the 
cross section within the respective \mtt
bin at NLO. This ratio is obtained from the \mtt distributions
computed in Sec.~\ref{subsec:mttbar}.  We observe that for all
parameter scenarios the S/B ratio of one or several of the four spin
observables outreaches the green band. Thus these spin observables
are, within the respective \mtt bin, more sensitive than the cross
section. Notice that in some cases the S/B ratio of the cross section
is very low and the green area is not (see for example the result for 
the high \mtt bin in Fig.~\ref{fig:spincorr_1a}) or only
barely visible (Fig.~\ref{fig:spincorr_2}).

For scenarios 1a--1c \Crr is the most sensitive of the four $P$- and
$CP$-even spin correlation observables.  The S/B ratio associated with
\Crr is also larger in magnitude than that of the cross section---in
case of scenario 1a almost by a factor of four in the lower \mtt bin.
In the higher \mtt bin of 400--460 GeV the spin correlations and
associated S/B ratios at NLO are smaller compared to those for the
lower \mtt bin.  In case of the observable \Crr the S/B ratios in the
higher \mtt bin of 400--460 GeV at NLO deviate significantly from
those at LO.
The reason for this large difference is the fact that the \mtt\ 
distribution for this observable has a zero at LO within this \mtt\ 
bin---that this, it receives positive and negative contributions that
(almost) cancel. This zero is shifted close to the bin boundary or
even outside the bin when taking NLO corrections into account. Because
the observable \Crr suffers from accidental (partial) cancellations at
leading order it is very sensitive to NLO corrections. This does not
signal a general breakdown of the perturbative expansion.

This circumstance should be seen as an artefact of the specific observable and is
associated to a large extend with the chosen $\mtt$ bin. Because of
this (accidentally) large sensitivity to higher-order corrections, we
observe large K-factors for \Crr in these specific cases:
$\text{K}=1.7$ for scenario 1a and $|\text{K}|>7$ for scenarios 1b and
1c. Moreover, the NLO corrections lie outside of the LO uncertainty
estimate. However, one may argue that the uncertainty estimate of the
LO predictions are in any case unreliable because the spin observables
studied here are defined as ratios where, for example, $\alpha_s$
cancels to leading order.  (The situation becomes even worse at LO
when ratios of ratios are considered.)  Alternatively, one can compute
a spin observable at NLO also without expanding the denominator and
then compute the ratio $R$ of the expanded and the unexpanded
version. This ratio may be viewed as indicative of the convergence of
the perturbative expansion.  For the observable \Crr evaluated in the
higher \mtt bin, we obtain $R=1.2$, $R=1.9$, and $R=1.5$ for scenario
1a, 1b, and 1c, respectively, whereas for the other three spin
observables we get $R\le1.1$. This shows again that in the \mtt bin of
400--460 GeV the observable \Crr is very sensitive to higher order
corrections.  It illustrates also the importance of the NLO
contributions to the spin observables.  This example shows that one
should be careful when using a specific spin observable in the search
for heavy Higgs effects in $\ttbar$ data.  Given a certain $\mtt$
bin-choice, one should use in an agnostic experimental analysis spin
observables that do not have a zero (in the above sense) at LO QCD.

As can be seen from \Figs{fig:spincorr_1a}--\ref{fig:spincorr_2} the
uncertainty bands of the values of a spin observable at LO and NLO due
to scale variations do not overlap in a number of cases. As stated
before this is mainly because the spin correlations are ratios (see
for example \Eq{eq:spinasymmetry} and \Eq{eq:SingleDiffDistribution}) 
in which the $\mur$ dependence cancels
at LO. The
leading-order scale variation thus underestimates the effect of
higher-order corrections and fails to give reliable uncertainty
estimates.

The values of the four $P$- and $CP$-even spin correlations and the
corresponding S/B ratios evaluated in the two \mtt bins displayed in
\Fig{fig:spincorr_1a} for scenario 1a are given in
\Tab{tab:spincorr_1a}.  In addition we list also the resulting values
when no cuts on \mtt are imposed.  Furthermore, the last column of
this table contains the cross section for the dileptonic $\ttbar$
decay channel summed over $\ell, \ell' = e,\mu$, both for the low and
high \mtt bin and inclusively, and the corresponding S/B ratios.  The
numbers substantiate what was already stated above: for the chosen
bins and inclusively, one or several of the above spin observables are
more sensitive to heavy Higgs-boson effects than the ratio of binned
cross sections. In the low \mtt bin \Cnn, \Crr and \D feature S/B
ratios larger than 6\%, i.e., a significant sensitivity to the
lighter of the two heavy Higgs bosons which is a scalar. Notice that
not only \Chel, \Cnn, and \D but also \Crr has a reliable perturbative
expansion in the lower \mtt bin.
 
The corresponding tables for the results presented in
\Fig{fig:spincorr_1b} and \Fig{fig:spincorr_1c} are given in
Appendix \ref{sec:spincorrelations_tables} 
(\Tab{tab:spincorr_1b} and \Tab{tab:spincorr_1c}). The predictions for the low
\mtt bin given in Tab.~\ref{tab:spincorr_1b} for scenario 1b with a
400 GeV pseudoscalar show that all four spin correlations and the
binned cross section have S/B$\gtrsim 10\%$.  In the lower \mtt bin of
scenario 1c only \Crr provides a S/B larger than that of the binned
cross section, namely $\sim 9\%$ (cf. Tab.~\ref{tab:spincorr_1c}).
Both for scenarios 1b and 1c the perturbative expansions of the spin
correlations in the low \mtt bin, in particular of \Crr, are reliable.

Scenarios 1a--1c feature strong Higgs-boson signals and experimental
analyses of the \mtt distribution might already be sensitive to these
effects. In scenario 2 the signal is much weaker which makes it
difficult to constrain this scenario with the \mtt distribution. As
can be seen from \Fig{fig:spincorr_2}---and from the corresponding
numbers in Tab.~\ref{tab:spincorr_2} of Appendix
\ref{sec:spincorrelations_tables}---the S/B ratio of the cross section
(green band) is very small in both \mtt bins, $\text{S/B}\lesssim
1\%$. As in scenarios 1a--1c we encounter here also the problem of
zero(s) of spin observables. In this case the observables \Chel and \D
have zeros in the \mtt bin 670--770 GeV at LO which are shifted to the
other \mtt bin 770--870 GeV at NLO. This leads to a high sensitivity
of \Chel and \D to NLO corrections which affects the robustness of the
prediction. The observables \Cnn and \Crr have a reliable perturbative
expansion in these \mtt bins.  However, taking the scale uncertainty
into account their sensitivity is not sufficient to constrain this
model. One way to remedy this is to evaluate \Chel, which is the most
sensitive of the four spin correlations in this scenario, within $500 \text{
GeV}\le\mtt\le750 \text{ GeV}$. This yields a S/B ratio of 4\% as
compared to the cross section S/B ratio of 0.7\% at NLO, and the
prediction is also more robust with respect to NLO corrections
($\text{K}=1.3$, $R=1.1$). The reason for having chosen the above bins
is that our prediction of the heavy Higgs effects at NLO is most
accurate within the Higgs resonance region.  Choosing instead an \mtt
range of 500--750 GeV will affect the accuracy of our approach. Thus,
a more careful investigation of the uncertainties of \Chel due to
these approximations would be necessary but is left for future work.

The spin correlations \Chel, \Crr, and \Cnn were recently computed,
inclusively in \mtt, at NLO QCD including weak interaction
corrections also in \cite{Bernreuther:2015yna}, and our corresponding
predictions in
Tabs.~\ref{tab:spincorr_1a},~\ref{tab:spincorr_1b},~\ref{tab:spincorr_1c}
and~\ref{tab:spincorr_2} agree with these results\footnote{The
  results presented here and those given in \Ref{Bernreuther:2015yna} 
  are not completely
  identical, because in \Ref{Bernreuther:2015yna} the scale choice $m_t/2\leq
  \mu\leq 2m_t$ was used.}.  The spin correlations \Chel and \Crr, \Cnn
were recently measured in dileptonic $\ttbar$ events, inclusively in
\mtt,  at the LHC(8 TeV) in 
\cite{Khachatryan:2016ngh,Aaboud:2016bit} and \cite{Aaboud:2016bit}, 
respectively. 
The results, corrected to parton level in the full 
phase space, agree  with respective SM predictions. At 8 TeV
center-of-mass 
energy the contributions of the heavy Higgs resonances of our
parameter scenarios to these observables are smaller in magnitude than
at 13 TeV, and we have checked that these contributions are in accord with the 
results of \cite{Khachatryan:2016ngh,Aaboud:2016bit} within the 
experimental uncertainties.

{\renewcommand{\arraystretch}{1.3}
\renewcommand{\tabcolsep}{0.2cm}
\begin{table}
\caption{$P$- and $CP$-even spin correlations and dileptonic cross section in scenario 1a for the LHC(13 TeV).}
\label{tab:spincorr_1a}\renewcommand{\arraystretch}{1.6}
\begin{tabular}{|c|c|c|c|c|c|c|c|}
\hline
\multicolumn{3}{|c|}{}& $C_{hel}$ & $C_{nn}$ & $C_{rr}$ & $D$ & $\sigma_{t\bar{t}}\times$BR [pb]\\
\hline
\multirow{6}{*}{\rotatebox[origin=c]{90}{\scriptsize $2m_t\le M_{t\bar{t}}\le 400\mbox{ GeV}$}} &
\multirow{3}{*}{\rotatebox[origin=c]{90}{LO}} & \scriptsize 2HDM+QCD &
$0.576_{-0.011}^{+0.011}$ & $0.483_{-0.005}^{+0.005}$ & $0.252_{-0.010}^{+0.010}$ & $-0.437_{-0.009}^{+0.009}$ & $3.70_{-0.68}^{+0.89}$\\
\cline{3-8}
& & \scriptsize QCD &
$0.562_{-0.011}^{+0.011}$ & $0.528_{-0.005}^{+0.005}$ & $0.289_{-0.011}^{+0.011}$ & $-0.459_{-0.009}^{+0.009}$ & $3.58_{-0.65}^{+0.85}$\\
\cline{3-8}
& & \scriptsize $\frac{\mbox{2HDM}}{\mbox{QCD}}$ [\%] &
$2.52_{-0.08}^{+0.08}$ & $-8.52_{-0.07}^{+0.06}$ & $-12.9_{-0.2}^{+0.2}$ & $-4.93_{-0.04}^{+0.03}$ & $3.35_{-0.05}^{+0.05}$\\
\cline{2-8}
& \multirow{3}{*}{\rotatebox[origin=c]{90}{NLO}} & \scriptsize 2HDM+QCDW &
$0.567_{-0.010}^{+0.008}$ & $0.462_{-0.008}^{+0.006}$ & $0.278_{-0.005}^{+0.005}$ & $-0.435_{-0.007}^{+0.007}$ & $6.30_{-0.76}^{+0.86}$\\
\cline{3-8}
& & \scriptsize QCDW &
$0.548_{-0.011}^{+0.009}$ & $0.522_{-0.005}^{+0.004}$ & $0.330_{-0.002}^{+0.003}$ & $-0.466_{-0.005}^{+0.006}$ & $6.06_{-0.73}^{+0.81}$\\
\cline{3-8}
& &\cellcolor{gray!50}\scriptsize $\frac{\mbox{2HDM}}{\mbox{QCDW}}$ [\%] &
\cellcolor{gray!50}$3.38_{-0.17}^{+0.28}$ & \cellcolor{gray!50}$-11.5_{-0.7}^{+0.5}$ & \cellcolor{gray!50}$-15.6_{-0.9}^{+0.7}$ & \cellcolor{gray!50}$-6.64_{-0.41}^{+0.33}$ & \cellcolor{gray!50}$3.97_{-0.18}^{+0.24}$\\
\hline\hline
\multirow{6}{*}{\rotatebox[origin=c]{90}{{\scriptsize $400\le M_{t\bar{t}}\le 460\mbox{ GeV}$}}} &
\multirow{3}{*}{\rotatebox[origin=c]{90}{LO}} & \scriptsize 2HDM+QCD &
$0.426_{-0.009}^{+0.009}$ & $0.374_{-0.002}^{+0.002}$ & $0.060_{-0.006}^{+0.006}$ & $-0.286_{-0.006}^{+0.006}$ & $4.49_{-0.86}^{+1.10}$\\
\cline{3-8}
& & \scriptsize QCD &
$0.432_{-0.009}^{+0.009}$ & $0.356_{-0.002}^{+0.002}$ & $0.045_{-0.006}^{+0.006}$ & $-0.278_{-0.006}^{+0.006}$ & $4.54_{-0.87}^{+1.10}$\\
\cline{3-8}
& & \scriptsize $\frac{\mbox{2HDM}}{\mbox{QCD}}$ [\%] &
$-1.43_{-0.04}^{+0.04}$ & $5.01_{-0.03}^{+0.03}$ & $32.1_{-3.0}^{+5.0}$ & $3.14_{-0.01}^{+0.01}$ & $-1.08_{-0.01}^{+0.01}$\\
\cline{2-8}
& \multirow{3}{*}{\rotatebox[origin=c]{90}{NLO}} & \scriptsize 2HDM+QCDW &
$0.430_{-0.008}^{+0.008}$ & $0.356_{-0.001}^{+0.001}$ & $0.103_{-0.004}^{+0.005}$ & $-0.296_{-0.001}^{+0.001}$ & $7.52_{-0.91}^{+0.98}$\\
\cline{3-8}
& & \scriptsize QCDW &
$0.426_{-0.007}^{+0.007}$ & $0.362_{-0.000}^{+0.001}$ & $0.107_{-0.003}^{+0.004}$ & $-0.298_{-0.001}^{+0.001}$ & $7.53_{-0.91}^{+0.98}$\\
\cline{3-8}
& &\cellcolor{gray!50}\scriptsize $\frac{\mbox{2HDM}}{\mbox{QCDW}}$ [\%] &
\cellcolor{gray!50}$0.959_{-0.150}^{+0.219}$ & \cellcolor{gray!50}$-1.76_{-0.47}^{+0.39}$ & \cellcolor{gray!50}$-3.34_{-1.30}^{+1.00}$ & \cellcolor{gray!50}$-0.651_{-0.222}^{+0.194}$ & \cellcolor{gray!50}$-0.017_{-0.012}^{+0.031}$\\
\hline\hline
\multirow{6}{*}{\rotatebox[origin=c]{90}{{\scriptsize incl. in $M_{t\bar{t}}$}}} &
\multirow{3}{*}{\rotatebox[origin=c]{90}{LO}} & \scriptsize 2HDM+QCD &
$0.298_{-0.004}^{+0.003}$ & $0.328_{-0.000}^{+0.000}$ & $-0.017_{-0.003}^{+0.002}$ & $-0.203_{-0.002}^{+0.002}$ & $18.7_{-4.0}^{+5.0}$\\
\cline{3-8}
& & \scriptsize QCD &
$0.297_{-0.004}^{+0.003}$ & $0.329_{-0.001}^{+0.000}$ & $-0.017_{-0.003}^{+0.002}$ & $-0.203_{-0.002}^{+0.002}$ & $18.6_{-4.0}^{+5.0}$\\
\cline{3-8}
& & \scriptsize $\frac{\mbox{2HDM}}{\mbox{QCD}}$ [\%] &
$0.313_{-0.031}^{+0.029}$ & $-0.277_{-0.019}^{+0.022}$ & $2.67_{-0.21}^{+0.19}$ & $-0.071_{-0.004}^{+0.005}$ & $0.133_{-0.011}^{+0.009}$\\
\cline{2-8}
& \multirow{3}{*}{\rotatebox[origin=c]{90}{NLO}} & \scriptsize 2HDM+QCDW &
$0.339_{-0.002}^{+0.004}$ & $0.317_{-0.002}^{+0.002}$ & $0.059_{-0.004}^{+0.006}$ & $-0.238_{-0.001}^{+0.000}$ & $30.5_{-4.0}^{+4.0}$\\
\cline{3-8}
& & \scriptsize QCDW &
$0.333_{-0.002}^{+0.003}$ & $0.327_{-0.001}^{+0.001}$ & $0.065_{-0.004}^{+0.006}$ & $-0.242_{-0.001}^{+0.000}$ & $30.3_{-4.0}^{+4.0}$\\
\cline{3-8}
& &\cellcolor{gray!50}\scriptsize $\frac{\mbox{2HDM}}{\mbox{QCDW}}$ [\%] &
\cellcolor{gray!50}$1.99_{-0.03}^{+0.03}$ & \cellcolor{gray!50}$-2.95_{-0.13}^{+0.10}$ & \cellcolor{gray!50}$-9.98_{-0.09}^{+0.17}$ & \cellcolor{gray!50}$-1.31_{-0.14}^{+0.11}$ & \cellcolor{gray!50}$0.631_{-0.027}^{+0.044}$\\
\hline
\end{tabular}

\end{table}

{\renewcommand{\arraystretch}{1.3}
\renewcommand{\tabcolsep}{0.2cm}
\begin{table}
\caption{The expectation value $\langle \Obs_{CP}\rangle$ in scenario
  1c for dileptonic $\ttbar$ events  at the LHC at 13 TeV.
  }
\label{tab:Ocp}
\begin{center}\renewcommand{\arraystretch}{1.6}
\begin{tabular}{ccc}
\hline\hline
$M_{t\bar{t}}$ {\small [GeV]} & {\small LO 2HDM} & {\small NLO 2HDM} \\
\hline
[$2m_t$,400] &
$-0.549_{-0.007}^{+0.007}$$\times 10^{-2}$ & $-0.824_{-0.029}^{+0.024}$$\times 10^{-2}$\\
$[400,460]$ &
$0.587_{-0.005}^{+0.005}$$\times 10^{-2}$ & $0.127_{-0.054}^{+0.062}$$\times 10^{-2}$\\
incl. &
$0.666_{-0.005}^{+0.005}$$\times 10^{-3}$ & $-0.814_{-0.020}^{+0.008}$$\times 10^{-3}$\\
\hline\hline
\end{tabular}

\end{center}
\end{table}}%

The observables \Chel, \Cnn, \Crr, and \D no not
provide any information about the $CP$ nature of the Higgs resonances.
The $P$- and $CP$-odd triple correlation $\OCP$ given in \Eq{eq:OCP} 
allows to search for (non-standard) CP violation in 
dileptonic $\ttbar$ events, in particular for effects induced by heavy
Higgs resonances that are $CP$ mixtures.
Among the four scenarios studied in this paper only the \CP-violating
scenario 1c leads to a non-vanishing  $\langle\Obs_{CP}\rangle$. 
The results for the expectation value 
$\langle \Obs_{CP}\rangle$ for dileptonic $\ttbar$ events at 13 TeV
in the low and high \mtt bin and inclusively in \mtt are 
presented in \Tab{tab:Ocp}.

The expectation value $\langle\Obs_{CP}\rangle$ is below the percent level 
 even in the bins around 400 GeV where
the $CP$-violating effect is caused by the resonant production of the
400 GeV Higgs resonance of indefinite parity.  The inclusive
expectation value is an order of magnitude smaller which is due to
partial cancellations of the two heavy Higgs boson contributions.
We note that $\langle\Obs_{CP}\rangle$ can also be obtained from the 
measurement of the difference of two off-diagonal correlations
cf. \Eq{eq:OCPvsC}.  
This difference of spin correlations was recently measured by ATLAS
for dileptonic $\ttbar$ events at 8 TeV.
At the parton level in the full phase-space the result 
$C_{nr}-C_{rn}=-0.006\pm 0.108$ was obtained \cite{Aaboud:2016bit}. 
We checked that the value of $9\langle\Obs_{CP}\rangle$
at 8 TeV in scenario 1c is in accord with this result within the
experimental uncertainty.

Table~\ref{tab:Bt_1c} contains our predictions, for scenarios 1a - 1c,
of the coefficient $B_t(k)$ that measures the longitudinal
polarization of the top quark in the helicity basis
(cf. \Eq{eq:SingleDiffDistribution}). The predictions apply to dileptonic and to
$\ell^+ + \mbox{jets}$ events.  A non-zero value of $B_t$, which
requires $P$-violating interactions, is generated in the SM by the
weak interactions. In scenarios 1a and 1b the neutral Higgs
interactions are $P$-conserving. Thus in these scenarios $B_t$ is zero
at LO and given at NLO by the results listed in column `QCDW' of
Tab.~\ref{tab:Bt_1c}---neglecting contributions from the charged
Higgs boson. They are very small because we assume $H^\pm$ to be very
heavy.  In scenario 1c the neutral Higgs bosons induce $P$-violating
effects already at LO and the resulting values of $B_t$ at NLO within
the SM + 2HDM differ from those of the SM.  Although the S/B ratio can
be significantly enhanced by \mtt cuts the effects remain below the
percent level and exhibit large scale uncertainties. In scenario 2 the
neutral Higgs interactions are $P$-conserving. Thus, the NLO
predictions given in column `QCDW' of Tab.~\ref{tab:Bt_1c} apply also
to this scenario, provided the same scale choices are made as in
scenarios 1a--1c.  Our inclusive QCDW prediction of $B_t$ given in
Tab.~\ref{tab:Bt_1c} agrees with the corresponding prediction of
\cite{Bernreuther:2015yna}. (Notice that our scale choices differ from
those of \cite{Bernreuther:2015yna}.) The polarization observable
$B_t$ was measured inclusively in \mtt at 8 TeV by the ATLAS and CMS
experiment \cite{Khachatryan:2016xws,Aaboud:2016bit}, with an
absolute experimental uncertainty of a few percent. 
We checked that our inclusive SM+2HDM predictions at 8 TeV are in
accord with these measurements.

{\renewcommand{\arraystretch}{1.6}
\renewcommand{\tabcolsep}{0.2cm}
\begin{table}
\caption{
 The coefficient $B_t(k)$, which measures the longitudinal
 polarization of the top quark in the helicity basis in the SM and in
 scenario 1c, for dileptonic and semileptonic $\ttbar$ events at the
 LHC(13 TeV).  In scenarios 1a and 1b \Bt receives no contributions
 from neutral Higgs bosons. In these scenarios \Bt is given by the
 QCDW value in the table.}
\label{tab:Bt_1c}
\begin{center}
\begin{tabular}{ccccccc}
\hline\hline
$M_{t\bar{t}}$ {\small [GeV]} && LO && & NLO &\\
\cline{3-3}\cline{5-7}
&& {\small 2HDM} && {\small 2HDM+QCDW} & {\small QCDW} & {\small $\frac{\mbox{2HDM}}{\mbox{QCDW}}$}\\
\hline
[$2m_t$,400] &
& $0.110_{-0.001}^{+0.001}$$\times 10^{-1}$&  & $0.492_{+0.014}^{+0.020}$$\times 10^{-2}$ & $0.169_{-0.098}^{+0.120}$$\times 10^{-2}$ & $1.91_{-1.20}^{+4.30}$\\
$[400,460]$ &
& $-0.769_{-0.007}^{+0.007}$$\times 10^{-2}$&  & $0.403_{-0.293}^{+0.313}$$\times 10^{-2}$ & $0.316_{-0.139}^{+0.174}$$\times 10^{-2}$ & $0.275_{-0.655}^{+0.187}$\\
incl. &
& $-0.600_{-0.026}^{+0.024}$$\times 10^{-3}$&  & $0.617_{-0.185}^{+0.227}$$\times 10^{-2}$ & $0.553_{-0.164}^{+0.209}$$\times 10^{-2}$ & $0.114_{-0.007}^{-0.006}$\\
\hline\hline
\end{tabular}

\end{center}
\end{table}}%

\begin{figure}[htbp]
\includegraphics[scale=1.3]{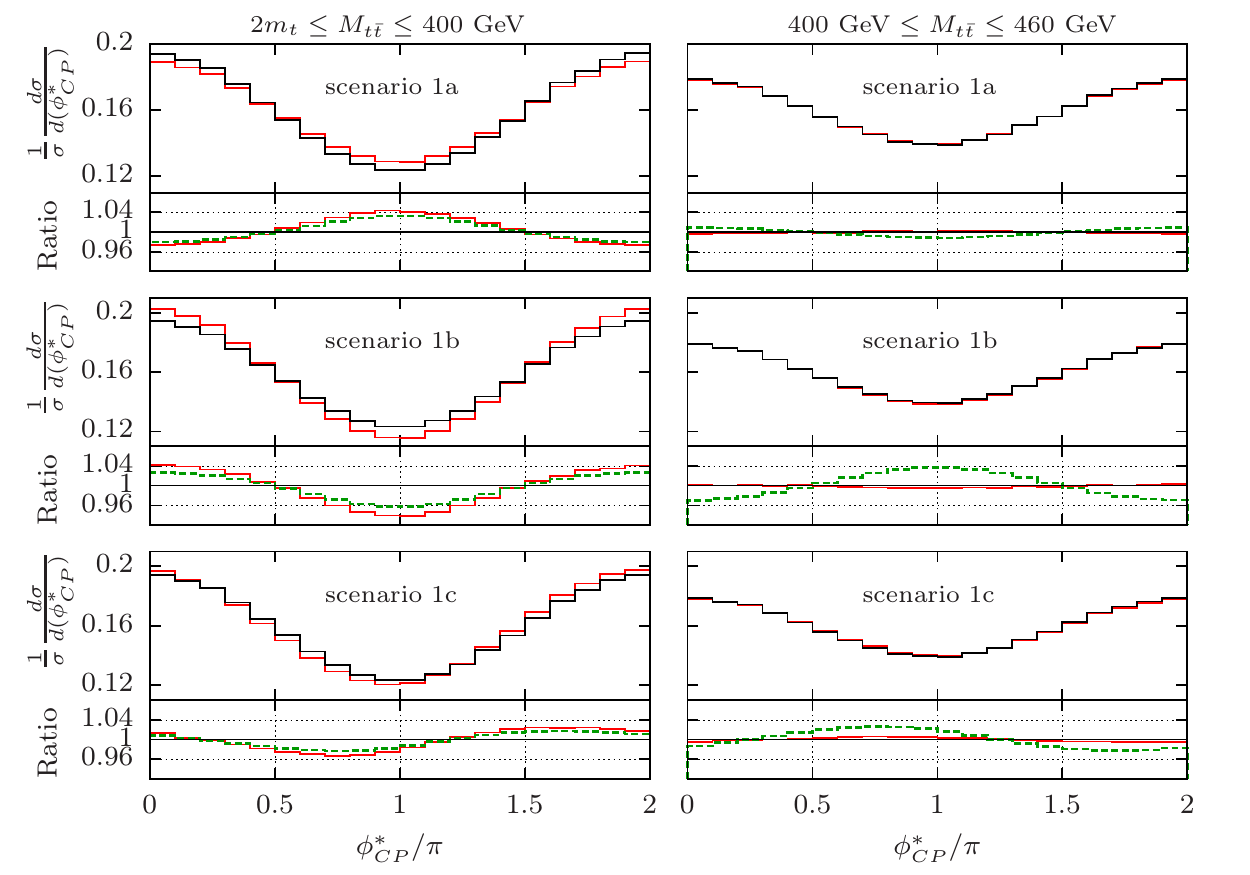}
\caption{Upper panels: The normalized distribution of \dphi for scenarios 1a--1c in two different \mtt bins.
 The NLO prediction in the SM and in the SM + 2HDM is displayed in
 black and in red, respectively. Lower panels: The curves in red
 (green) are the ratios (SM+2HDM)/SM at NLO (LO).  }
\label{fig:dphi}
\end{figure}

Finally we analyze the normalized leptonic azimuthal-angle difference
\dphi defined in Sec.~\ref{subsec:dphi}.  Observables of this type
have the potential to yield information about the $CP$ properties of
the Higgs bosons \cite{Baumgart:2011wk,Barger:2011pu,Berge:2014sra}.  We
compute the distribution of \dphi for scenarios 1a - 1c at NLO in the
SM and in the SM + 2HDM in the low and high \mtt bins that were
already chosen above for evaluating the other spin observables. These
scenarios were chosen on purpose such that they contain a scalar (1a),
a pseudoscalar (1b), or a \CP-mixed (1c) Higgs boson of the same mass (400
GeV). Our aim is to investigate the discriminating power of \dphi with
respect to the $CP$ nature of this boson at NLO.  (An observable
similar to \dphi with a different phase convention was analyzed at LO
in Ref.~\cite{Barger:2011pu}, including the interference with the QCD
background.)
 
Our results for the normalized distribution of \dphi in scenarios
1a--1c are displayed in the upper panels of \Fig{fig:dphi}, on the left
(right) for the low (high) \mtt bin.  The distributions in black and
red are the predictions within the SM and the SM + 2HDM, respectively,
at NLO. The effect of scale variations is very small and not visible
in these plots. In the lower panels the ratios R=(SM+2HDM)/SM are
plotted. Here the green-dashed and red-solid lines refer to the LO and NLO
predictions, respectively. These ratios have a trigonometric
function-like shape. The ratios R in the low \mtt bin have some
sensitivity to the CP nature of the 400 GeV Higgs boson. While the
maximum (minimum) of R is located at $\dphi=\pi$ $(\dphi=0,2\pi)$ if
the 400 GeV Higgs boson is a scalar, it is shifted to $\dphi=0, 2\pi$
$(\dphi=\pi)$ if this boson is a pseudoscalar. If the 400 GeV Higgs
boson is a $CP$ mixture as in scenario 1c the maximum and minimum of R
are located at $\dphi\neq 0, \pi, 2\pi$. The locations depend on the
relative strengths and phases of the scalar and pseudoscalar Yukawa
couplings of the 400 GeV Higgs boson to top quarks.  In contrast to
the tau-pair decay channel of a Higgs boson studied in
\Ref{Berge:2014sra}, the background is not flat in \dphi and the
signal-background interference is not negligible in the case at
hand. The large non-resonant SM background contribution leads to a
rather small signal-to-background ratio of $\text{S/B}\lesssim 6
\%$. We recall that in these scenarios the 400 GeV Higgs signal is
significant in the \mtt distribution and in spin observables studied
above.  We therefore expect even lower S/B ratios for weaker Higgs
signals, which poses an even bigger challenge to determine the Higgs
\CP properties with this observable. The NLO corrections enhance the
S/B ratio only slightly.

%
\section{Summary and conclusions}
\label{sec:summary}

We considered, in continuation of our previous work
\cite{Bernreuther:2015fts}, the production of top-quark pairs at the
LHC (13 TeV) and their subsequent decay to dileptonic final states and
analyzed, within the type-II 2HDM extension of the SM, the sensitivity
of top-spin dependent observables to the resonant production of heavy
Higgs bosons.  NLO QCD corrections to the Higgs-boson signal in the
$\ttbar$ channel and NLO QCD and weak interaction corrections to the
non-resonant $\ttbar$ background were taken into account, including
the signal-background interference at NLO. We determined, for four
different 2HDM parameter scenarios, a number of leptonic angular
correlations and distributions that correspond to \ttbar spin
correlations and to the longitudinal top-quark polarization. We
computed also the $\ttbar$ invariant mass distribution, which is the
basic observable in the search for heavy Higgs bosons in \ttbar
events, in order to assess the gain in sensitivity if such a search is
accompanied by top-spin dependent observables. Our analysis shows that
$\ttbar$ spin correlations have the potential to substantially
increase the sensitivity of the \ttbar channel to heavy Higgs
resonances if they are evaluated in judiciously chosen \mtt bins. NLO
corrections are, needless to say, important for assessing whether a
specific observable has a robust perturbative expansion within a
chosen \mtt bin and for estimating the uncertainties due to scale
variations.
  
The observable \Bt that corresponds to the longitudinal top-quark
polarization and the triple product correlation \OCP analyzed above
are sensitive to $P$-violating, respectively $P$- and $CP$-violating
interactions. We studied the effects of heavy Higgs resonances of
undefined $CP$ parity on these observables for a $CP$-violating 2HDM
scenario. We found that the effects are small, even if these
observables are evaluated in appropriate \mtt bins. Our studies
indicate that measurement of these (dimensionless) observables with
a precision of $\sim 10^{-3}$ would be required in order to reach
meaningful sensitivities to $P$- and $CP$-violating heavy
Higgs-boson effects.
 
Furthermore, we studied the potential of the lepton azimuthal angle
distribution $d\sigma/d\dphi$ for pinning down the \CP properties of a
heavy Higgs resonance in the dileptonic \ttbar events.  This
observable allows to construct a ratio whose shape shows significant
differences with respect to effects of Higgs bosons of different \CP
nature. Yet the signal-to-background ratio is only about
$\text{S/B}\lesssim 6\%$ because of the large SM background and the
non-negligible signal-background interference.
 
Our detailed studies show that including top-spin dependent
 observables in the toolkit for the search for heavy Higgs-boson
 resonances in the \ttbar channel can significantly enhance the
 sensitivity of these explorations.  Although the results of this
 paper were obtained for specific parameter settings within the 2HDM,
 we believe that this conclusion remains valid also for other SM
 extensions that predict heavy neutral Higgs bosons with unsuppressed
 couplings to top quarks. We showed also that inclusion of NLO QCD
 corrections is important in this type of analysis. In particular, we
 illustrated with a concrete example that a naive K-factor approach is
 in general not sufficient for obtaining reliable predictions.

%

\subsubsection*{Acknowledgments}

We thank C. Mellein for collaboration at an early stage of this project.
P. Galler  was supported by Deutsche Forschungsgemeinschaft through
Graduiertenkolleg Grant No. GRK 1504. The work  of Z.-G. Si was supported by
National Natural Science Foundation of China and by Natural Science Foundation
of Shandong Province. 

%
\appendix

\section{Results for the $P$- and $CP$-even spin correlations and cross section in scenarios 1b, 1c, and 2}
\label{sec:spincorrelations_tables}

We present the results displayed in
Figs.~\ref{fig:spincorr_1b},~\ref{fig:spincorr_1c},
and~\ref{fig:spincorr_2} for the parameter scenarios 1b, 1c, and 2,
respectively, in the tables of this appendix.  They contain the
values of the four $P$- and $CP$-even spin correlations and the
corresponding S/B ratios evaluated in two \mtt bins and, in addition,
in the whole \mtt range.  Moreover, the cross section for the
dileptonic $\ttbar$ decay channel summed over $\ell, \ell' = e,\mu$
is given in the last column of this table, both for the low and high
\mtt bin and inclusively, and the corresponding S/B ratios.  These
results are discussed in Sec.~\ref{subsec:spin}.

{\renewcommand{\arraystretch}{1.3}
\renewcommand{\tabcolsep}{0.2cm}
\begin{table}[htbp]
  \caption{$P$- and $CP$-even spin correlations and dileptonic cross
    section in scenario 1b for the LHC(13 TeV).}
  \label{tab:spincorr_1b}
  \begin{tabular}{|c|c|c|c|c|c|c|c|}
\hline
\multicolumn{3}{|c|}{}& $C_{hel}$ & $C_{nn}$ & $C_{rr}$ & $D$ & $\sigma_{t\bar{t}}\times$BR [pb]\\
\hline
\multirow{6}{*}{\rotatebox[origin=c]{90}{\scriptsize $2m_t\le M_{t\bar{t}}\le 400\mbox{ GeV}$}} &
\multirow{3}{*}{\rotatebox[origin=c]{90}{LO}} & \scriptsize 2HDM+QCD &
$0.599_{-0.011}^{+0.011}$ & $0.567_{-0.005}^{+0.005}$ & $0.349_{-0.010}^{+0.011}$ & $-0.505_{-0.009}^{+0.009}$ & $3.91_{-0.72}^{+0.94}$\\
\cline{3-8}
& & \scriptsize QCD &
$0.562_{-0.011}^{+0.011}$ & $0.528_{-0.005}^{+0.005}$ & $0.289_{-0.011}^{+0.011}$ & $-0.459_{-0.009}^{+0.009}$ & $3.58_{-0.65}^{+0.85}$\\
\cline{3-8}
& & \scriptsize $\frac{\mbox{2HDM}}{\mbox{QCD}}$ [\%] &
$6.61_{-0.22}^{+0.23}$ & $7.51_{-0.08}^{+0.08}$ & $20.7_{-0.8}^{+0.9}$ & $9.92_{-0.25}^{+0.26}$ & $9.28_{-0.11}^{+0.12}$\\
\cline{2-8}
& \multirow{3}{*}{\rotatebox[origin=c]{90}{NLO}} & \scriptsize 2HDM+QCDW &
$0.605_{-0.009}^{+0.008}$ & $0.582_{-0.003}^{+0.003}$ & $0.416_{+0.000}^{+0.001}$ & $-0.534_{-0.004}^{+0.004}$ & $6.81_{-0.84}^{+0.95}$\\
\cline{3-8}
& & \scriptsize QCDW &
$0.548_{-0.011}^{+0.009}$ & $0.522_{-0.005}^{+0.004}$ & $0.330_{-0.002}^{+0.003}$ & $-0.466_{-0.005}^{+0.006}$ & $6.06_{-0.73}^{+0.81}$\\
\cline{3-8}
& &\cellcolor{gray!50}\scriptsize $\frac{\mbox{2HDM}}{\mbox{QCDW}}$ [\%] &
\cellcolor{gray!50}$10.3_{-0.4}^{+0.6}$ & \cellcolor{gray!50}$11.4_{-0.3}^{+0.5}$ & \cellcolor{gray!50}$26.2_{-0.7}^{+0.9}$ & \cellcolor{gray!50}$14.5_{-0.5}^{+0.7}$ & \cellcolor{gray!50}$12.3_{-0.5}^{+0.5}$\\
\hline\hline
\multirow{6}{*}{\rotatebox[origin=c]{90}{{\scriptsize $400\le M_{t\bar{t}}\le 460\mbox{ GeV}$}}} &
\multirow{3}{*}{\rotatebox[origin=c]{90}{LO}} & \scriptsize 2HDM+QCD &
$0.396_{-0.009}^{+0.009}$ & $0.315_{-0.002}^{+0.002}$ & $-0.016_{-0.006}^{+0.006}$ & $-0.232_{-0.006}^{+0.006}$ & $4.28_{-0.81}^{+1.10}$\\
\cline{3-8}
& & \scriptsize QCD &
$0.432_{-0.009}^{+0.009}$ & $0.356_{-0.002}^{+0.002}$ & $0.045_{-0.006}^{+0.006}$ & $-0.278_{-0.006}^{+0.006}$ & $4.54_{-0.87}^{+1.10}$\\
\cline{3-8}
& & \scriptsize $\frac{\mbox{2HDM}}{\mbox{QCD}}$ [\%] &
$-8.20_{-0.23}^{+0.22}$ & $-11.6_{-0.0}^{0.0}$ & $-134_{-21}^{+16}$ & $-16.5_{-0.3}^{+0.3}$ & $-5.87_{-0.05}^{+0.05}$\\
\cline{2-8}
& \multirow{3}{*}{\rotatebox[origin=c]{90}{NLO}} & \scriptsize 2HDM+QCDW &
$0.429_{-0.013}^{+0.012}$ & $0.366_{-0.005}^{+0.005}$ & $0.116_{-0.004}^{+0.004}$ & $-0.304_{-0.007}^{+0.008}$ & $7.36_{-0.88}^{+0.96}$\\
\cline{3-8}
& & \scriptsize QCDW &
$0.426_{-0.007}^{+0.007}$ & $0.362_{-0.000}^{+0.001}$ & $0.107_{-0.003}^{+0.004}$ & $-0.298_{-0.001}^{+0.001}$ & $7.53_{-0.91}^{+0.98}$\\
\cline{3-8}
& &\cellcolor{gray!50}\scriptsize $\frac{\mbox{2HDM}}{\mbox{QCDW}}$ [\%] &
\cellcolor{gray!50}$0.844_{-1.370}^{+1.150}$ & \cellcolor{gray!50}$1.06_{-1.50}^{+1.30}$ & \cellcolor{gray!50}$8.39_{-7.20}^{+7.00}$ & \cellcolor{gray!50}$1.83_{-2.10}^{+1.90}$ & \cellcolor{gray!50}$-2.25_{-0.04}^{+0.09}$\\
\hline\hline
\multirow{6}{*}{\rotatebox[origin=c]{90}{{\scriptsize incl. in $M_{t\bar{t}}$}}} &
\multirow{3}{*}{\rotatebox[origin=c]{90}{LO}} & \scriptsize 2HDM+QCD &
$0.297_{-0.004}^{+0.003}$ & $0.329_{-0.001}^{+0.000}$ & $-0.017_{-0.002}^{+0.002}$ & $-0.203_{-0.002}^{+0.002}$ & $18.7_{-4.0}^{+5.0}$\\
\cline{3-8}
& & \scriptsize QCD &
$0.297_{-0.004}^{+0.003}$ & $0.329_{-0.001}^{+0.000}$ & $-0.017_{-0.003}^{+0.002}$ & $-0.203_{-0.002}^{+0.002}$ & $18.6_{-4.0}^{+5.0}$\\
\cline{3-8}
& & \scriptsize $\frac{\mbox{2HDM}}{\mbox{QCD}}$ [\%] &
$0.043_{-0.045}^{+0.042}$ & $-0.032_{-0.029}^{+0.026}$ & $0.261_{-0.781}^{+1.140}$ & $-0.004_{-0.064}^{+0.059}$ & $0.018_{-0.019}^{+0.017}$\\
\cline{2-8}
& \multirow{3}{*}{\rotatebox[origin=c]{90}{NLO}} & \scriptsize 2HDM+QCDW &
$0.351_{-0.003}^{+0.004}$ & $0.343_{-0.002}^{+0.002}$ & $0.090_{-0.004}^{+0.006}$ & $-0.261_{-0.001}^{+0.000}$ & $30.8_{-4.0}^{+4.0}$\\
\cline{3-8}
& & \scriptsize QCDW &
$0.333_{-0.002}^{+0.003}$ & $0.327_{-0.001}^{+0.001}$ & $0.065_{-0.004}^{+0.006}$ & $-0.242_{-0.001}^{+0.000}$ & $30.3_{-4.0}^{+4.0}$\\
\cline{3-8}
& &\cellcolor{gray!50}\scriptsize $\frac{\mbox{2HDM}}{\mbox{QCDW}}$ [\%] &
\cellcolor{gray!50}$5.41_{-0.17}^{+0.17}$ & \cellcolor{gray!50}$4.97_{-0.08}^{+0.10}$ & \cellcolor{gray!50}$38.4_{-4.0}^{+4.0}$ & \cellcolor{gray!50}$8.18_{-0.26}^{+0.26}$ & \cellcolor{gray!50}$1.57_{-0.07}^{+0.12}$\\
\hline
\end{tabular}

\end{table}}%
{\renewcommand{\arraystretch}{1.3}
\renewcommand{\tabcolsep}{0.2cm}
\begin{table}[htbp]
\caption{$P$- and $CP$-even spin correlations and dileptonic cross section in scenario 1c for the LHC(13 TeV).}
\label{tab:spincorr_1c}\renewcommand{\arraystretch}{1.6}
\begin{tabular}{|c|c|c|c|c|c|c|c|}
\hline
\multicolumn{3}{|c|}{}& $C_{hel}$ & $C_{nn}$ & $C_{rr}$ & $D$ & $\sigma_{t\bar{t}}\times$BR [pb]\\
\hline
\multirow{6}{*}{\rotatebox[origin=c]{90}{\scriptsize $2m_t\le M_{t\bar{t}}\le 400\mbox{ GeV}$}} &
\multirow{3}{*}{\rotatebox[origin=c]{90}{LO}} & \scriptsize 2HDM+QCD &
$0.582_{-0.011}^{+0.011}$ & $0.539_{-0.005}^{+0.005}$ & $0.311_{-0.010}^{+0.011}$ & $-0.477_{-0.009}^{+0.009}$ & $3.75_{-0.68}^{+0.90}$\\
\cline{3-8}
& & \scriptsize QCD &
$0.562_{-0.011}^{+0.011}$ & $0.528_{-0.005}^{+0.005}$ & $0.289_{-0.011}^{+0.011}$ & $-0.459_{-0.009}^{+0.009}$ & $3.58_{-0.65}^{+0.85}$\\
\cline{3-8}
& & \scriptsize $\frac{\mbox{2HDM}}{\mbox{QCD}}$ [\%] &
$3.51_{-0.12}^{+0.12}$ & $2.21_{-0.05}^{+0.05}$ & $7.78_{-0.36}^{+0.38}$ & $3.91_{-0.12}^{+0.13}$ & $4.73_{-0.06}^{+0.06}$\\
\cline{2-8}
& \multirow{3}{*}{\rotatebox[origin=c]{90}{NLO}} & \scriptsize 2HDM+QCDW &
$0.577_{-0.010}^{+0.008}$ & $0.539_{-0.005}^{+0.004}$ & $0.360_{-0.002}^{+0.002}$ & $-0.492_{-0.005}^{+0.005}$ & $6.42_{-0.78}^{+0.87}$\\
\cline{3-8}
& & \scriptsize QCDW &
$0.548_{-0.011}^{+0.009}$ & $0.522_{-0.005}^{+0.004}$ & $0.330_{-0.002}^{+0.003}$ & $-0.466_{-0.005}^{+0.006}$ & $6.06_{-0.73}^{+0.81}$\\
\cline{3-8}
& &\cellcolor{gray!50}\scriptsize $\frac{\mbox{2HDM}}{\mbox{QCDW}}$ [\%] &
\cellcolor{gray!50}$5.18_{-0.29}^{+0.29}$ & \cellcolor{gray!50}$3.32_{-0.07}^{+0.10}$ & \cellcolor{gray!50}$9.24_{-0.21}^{+0.22}$ & \cellcolor{gray!50}$5.45_{-0.18}^{+0.21}$ & \cellcolor{gray!50}$5.93_{-0.20}^{+0.27}$\\
\hline\hline
\multirow{6}{*}{\rotatebox[origin=c]{90}{{\scriptsize $400\le M_{t\bar{t}}\le 460\mbox{ GeV}$}}} &
\multirow{3}{*}{\rotatebox[origin=c]{90}{LO}} & \scriptsize 2HDM+QCD &
$0.408_{-0.009}^{+0.009}$ & $0.337_{-0.002}^{+0.002}$ & $0.013_{-0.006}^{+0.006}$ & $-0.253_{-0.006}^{+0.006}$ & $4.36_{-0.83}^{+1.10}$\\
\cline{3-8}
& & \scriptsize QCD &
$0.432_{-0.009}^{+0.009}$ & $0.356_{-0.002}^{+0.002}$ & $0.045_{-0.006}^{+0.006}$ & $-0.278_{-0.006}^{+0.006}$ & $4.54_{-0.87}^{+1.10}$\\
\cline{3-8}
& & \scriptsize $\frac{\mbox{2HDM}}{\mbox{QCD}}$ [\%] &
$-5.41_{-0.15}^{+0.15}$ & $-5.47_{-0.01}^{+0.01}$ & $-71.7_{-12.0}^{+9.0}$ & $-9.02_{-0.20}^{+0.19}$ & $-3.95_{-0.03}^{+0.03}$\\
\cline{2-8}
& \multirow{3}{*}{\rotatebox[origin=c]{90}{NLO}} & \scriptsize 2HDM+QCDW &
$0.421_{-0.010}^{+0.010}$ & $0.356_{-0.002}^{+0.003}$ & $0.100_{+0.000}^{+0.001}$ & $-0.293_{-0.004}^{+0.004}$ & $7.36_{-0.88}^{+0.95}$\\
\cline{3-8}
& & \scriptsize QCDW &
$0.426_{-0.007}^{+0.007}$ & $0.362_{-0.000}^{+0.001}$ & $0.107_{-0.003}^{+0.004}$ & $-0.298_{-0.001}^{+0.001}$ & $7.53_{-0.91}^{+0.98}$\\
\cline{3-8}
& &\cellcolor{gray!50}\scriptsize $\frac{\mbox{2HDM}}{\mbox{QCDW}}$ [\%] &
\cellcolor{gray!50}$-1.15_{-0.74}^{+0.67}$ & \cellcolor{gray!50}$-1.65_{-0.61}^{+0.55}$ & \cellcolor{gray!50}$-6.02_{-3.00}^{+3.00}$ & \cellcolor{gray!50}$-1.91_{-1.00}^{+0.88}$ & \cellcolor{gray!50}$-2.15_{-0.09}^{+0.08}$\\
\hline\hline
\multirow{6}{*}{\rotatebox[origin=c]{90}{{\scriptsize incl. in $M_{t\bar{t}}$}}} &
\multirow{3}{*}{\rotatebox[origin=c]{90}{LO}} & \scriptsize 2HDM+QCD &
$0.295_{-0.004}^{+0.003}$ & $0.328_{-0.001}^{+0.000}$ & $-0.019_{-0.003}^{+0.002}$ & $-0.201_{-0.002}^{+0.002}$ & $18.6_{-4.0}^{+5.0}$\\
\cline{3-8}
& & \scriptsize QCD &
$0.297_{-0.004}^{+0.003}$ & $0.329_{-0.001}^{+0.000}$ & $-0.017_{-0.003}^{+0.002}$ & $-0.203_{-0.002}^{+0.002}$ & $18.6_{-4.0}^{+5.0}$\\
\cline{3-8}
& & \scriptsize $\frac{\mbox{2HDM}}{\mbox{QCD}}$ [\%] &
$-0.735_{-0.013}^{+0.009}$ & $-0.294_{-0.007}^{+0.006}$ & $12.0_{-2.0}^{+2.0}$ & $-0.853_{-0.015}^{+0.011}$ & $-0.310_{-0.011}^{+0.010}$\\
\cline{2-8}
& \multirow{3}{*}{\rotatebox[origin=c]{90}{NLO}} & \scriptsize 2HDM+QCDW &
$0.339_{-0.003}^{+0.004}$ & $0.331_{-0.001}^{+0.002}$ & $0.073_{-0.004}^{+0.006}$ & $-0.248_{-0.001}^{+0.000}$ & $30.5_{-4.0}^{+4.0}$\\
\cline{3-8}
& & \scriptsize QCDW &
$0.333_{-0.002}^{+0.003}$ & $0.327_{-0.001}^{+0.001}$ & $0.065_{-0.004}^{+0.006}$ & $-0.242_{-0.001}^{+0.000}$ & $30.3_{-4.0}^{+4.0}$\\
\cline{3-8}
& &\cellcolor{gray!50}\scriptsize $\frac{\mbox{2HDM}}{\mbox{QCDW}}$ [\%] &
\cellcolor{gray!50}$2.01_{-0.17}^{+0.16}$ & \cellcolor{gray!50}$1.21_{-0.09}^{+0.09}$ & \cellcolor{gray!50}$11.4_{-2.0}^{+2.0}$ & \cellcolor{gray!50}$2.49_{-0.21}^{+0.20}$ & \cellcolor{gray!50}$0.442_{-0.017}^{+0.033}$\\
\hline
\end{tabular}

\end{table}}%
{\renewcommand{\arraystretch}{1.3}
\renewcommand{\tabcolsep}{0.2cm}
\begin{table}[htbp]
\caption{$P$- and $CP$-even spin correlations and dileptonic cross section in scenario 2 for the LHC(13 TeV).}
\label{tab:spincorr_2}\renewcommand{\arraystretch}{1.6}
\begin{tabular}{|c|c|c|c|c|c|c|c|}
\hline
\multicolumn{3}{|c|}{}& $C_{hel}$ & $C_{nn}$ & $C_{rr}$ & $D$ & $\sigma_{t\bar{t}}\times$BR [pb]\\
\hline
\multirow{6}{*}{\rotatebox[origin=c]{90}{\scriptsize $670\le M_{t\bar{t}}\le 770\mbox{ GeV}$}} &
\multirow{3}{*}{\rotatebox[origin=c]{90}{LO}} & \scriptsize 2HDM+QCD &
$0.012_{-0.006}^{+0.006}$ & $0.217_{-0.001}^{+0.001}$ & $-0.217_{-0.003}^{+0.003}$ & $-0.004_{-0.003}^{+0.003}$ & $1.16_{-0.25}^{+0.35}$\\
\cline{3-8}
& & \scriptsize QCD &
$0.021_{-0.006}^{+0.006}$ & $0.226_{-0.001}^{+0.001}$ & $-0.204_{-0.003}^{+0.003}$ & $-0.014_{-0.003}^{+0.003}$ & $1.17_{-0.25}^{+0.35}$\\
\cline{3-8}
& & \scriptsize $\frac{\mbox{2HDM}}{\mbox{QCD}}$ [\%] &
$-42.9_{-18.0}^{+10.0}$ & $-3.95_{-0.06}^{+0.05}$ & $6.28_{-0.12}^{+0.14}$ & $-72.4_{-16.0}^{+11.0}$ & $-0.899_{-0.011}^{+0.010}$\\
\cline{2-8}
& \multirow{3}{*}{\rotatebox[origin=c]{90}{NLO}} & \scriptsize 2HDM+QCDW &
$0.125_{-0.012}^{+0.011}$ & $0.198_{-0.004}^{+0.004}$ & $-0.103_{-0.001}^{+0.000}$ & $-0.072_{-0.005}^{+0.005}$ & $1.91_{-0.25}^{+0.25}$\\
\cline{3-8}
& & \scriptsize QCDW &
$0.103_{-0.000}^{+0.001}$ & $0.201_{-0.002}^{+0.002}$ & $-0.110_{-0.007}^{+0.008}$ & $-0.063_{-0.003}^{+0.002}$ & $1.90_{-0.25}^{+0.26}$\\
\cline{3-8}
& &\cellcolor{gray!50}\scriptsize $\frac{\mbox{2HDM}}{\mbox{QCDW}}$ [\%] &
\cellcolor{gray!50}$21.8_{-12.0}^{+11.0}$ & \cellcolor{gray!50}$-1.34_{-1.00}^{+0.81}$ & \cellcolor{gray!50}$-6.95_{-5.10}^{+6.80}$ & \cellcolor{gray!50}$14.3_{-12.0}^{+11.0}$ & \cellcolor{gray!50}$0.953_{-0.628}^{+0.433}$\\
\hline\hline
\multirow{6}{*}{\rotatebox[origin=c]{90}{{\scriptsize $770\le M_{t\bar{t}}\le 870\mbox{ GeV}$}}} &
\multirow{3}{*}{\rotatebox[origin=c]{90}{LO}} & \scriptsize 2HDM+QCD &
$-0.083_{-0.006}^{+0.006}$ & $0.232_{-0.001}^{+0.001}$ & $-0.215_{-0.003}^{+0.003}$ & $0.022_{-0.002}^{+0.002}$ & $0.622_{-0.139}^{+0.194}$\\
\cline{3-8}
& & \scriptsize QCD &
$-0.067_{-0.006}^{+0.006}$ & $0.223_{-0.001}^{+0.001}$ & $-0.217_{-0.003}^{+0.003}$ & $0.020_{-0.002}^{+0.002}$ & $0.631_{-0.141}^{+0.197}$\\
\cline{3-8}
& & \scriptsize $\frac{\mbox{2HDM}}{\mbox{QCD}}$ [\%] &
$25.2_{-2.0}^{+2.0}$ & $4.04_{-0.04}^{+0.04}$ & $-0.956_{-0.039}^{+0.036}$ & $9.42_{-0.88}^{+1.10}$ & $-1.55_{-0.01}^{+0.01}$\\
\cline{2-8}
& \multirow{3}{*}{\rotatebox[origin=c]{90}{NLO}} & \scriptsize 2HDM+QCDW &
$0.032_{-0.007}^{+0.009}$ & $0.185_{-0.001}^{0.000}$ & $-0.117_{-0.010}^{+0.015}$ & $-0.034_{-0.007}^{+0.007}$ & $1.01_{-0.14}^{+0.15}$\\
\cline{3-8}
& & \scriptsize QCDW &
$0.041_{-0.003}^{+0.004}$ & $0.185_{-0.004}^{+0.002}$ & $-0.115_{-0.006}^{+0.010}$ & $-0.037_{-0.003}^{+0.003}$ & $1.02_{-0.14}^{+0.15}$\\
\cline{3-8}
& &\cellcolor{gray!50}\scriptsize $\frac{\mbox{2HDM}}{\mbox{QCDW}}$ [\%] &
\cellcolor{gray!50}$-21.7_{-13.0}^{+13.0}$ & \cellcolor{gray!50}$0.279_{-1.250}^{+1.460}$ & \cellcolor{gray!50}$1.52_{-4.60}^{+3.40}$ & \cellcolor{gray!50}$-9.13_{-12.00}^{+11.00}$ & \cellcolor{gray!50}$-1.20_{-0.26}^{+0.37}$\\
\hline\hline
\multirow{6}{*}{\rotatebox[origin=c]{90}{{\scriptsize incl. in $M_{t\bar{t}}$}}} &
\multirow{3}{*}{\rotatebox[origin=c]{90}{LO}} & \scriptsize 2HDM+QCD &
$0.298_{-0.004}^{+0.003}$ & $0.329_{-0.000}^{+0.000}$ & $-0.016_{-0.003}^{+0.002}$ & $-0.204_{-0.002}^{+0.002}$ & $17.8_{-4.0}^{+5.0}$\\
\cline{3-8}
& & \scriptsize QCD &
$0.296_{-0.004}^{+0.004}$ & $0.329_{-0.000}^{+0.000}$ & $-0.018_{-0.003}^{+0.002}$ & $-0.203_{-0.002}^{+0.002}$ & $17.7_{-4.0}^{+5.0}$\\
\cline{3-8}
& & \scriptsize $\frac{\mbox{2HDM}}{\mbox{QCD}}$ [\%] &
$0.589_{-0.019}^{+0.019}$ & $0.061_{-0.004}^{+0.003}$ & $-6.05_{-0.71}^{+0.66}$ & $0.494_{-0.018}^{+0.017}$ & $0.249_{-0.004}^{+0.003}$\\
\cline{2-8}
& \multirow{3}{*}{\rotatebox[origin=c]{90}{NLO}} & \scriptsize 2HDM+QCDW &
$0.337_{-0.003}^{+0.004}$ & $0.328_{-0.001}^{+0.002}$ & $0.066_{-0.004}^{+0.005}$ & $-0.244_{-0.001}^{+0.000}$ & $29.6_{-4.0}^{+4.0}$\\
\cline{3-8}
& & \scriptsize QCDW &
$0.333_{-0.002}^{+0.004}$ & $0.327_{-0.001}^{+0.002}$ & $0.064_{-0.004}^{+0.006}$ & $-0.242_{-0.001}^{+0.000}$ & $29.5_{-4.0}^{+4.0}$\\
\cline{3-8}
& &\cellcolor{gray!50}\scriptsize $\frac{\mbox{2HDM}}{\mbox{QCDW}}$ [\%] &
\cellcolor{gray!50}$1.08_{-0.12}^{+0.10}$ & \cellcolor{gray!50}$0.159_{-0.026}^{+0.022}$ & \cellcolor{gray!50}$3.38_{-0.74}^{+0.69}$ & \cellcolor{gray!50}$0.867_{-0.119}^{+0.100}$ & \cellcolor{gray!50}$0.421_{-0.023}^{+0.016}$\\
\hline
\end{tabular}

\end{table}}%

%
\clearpage
\bibliographystyle{utphysmod}
\bibliography{spinreferences}

\end{document}